\documentclass[sigconf, nonacm]{acmart}

\newcommand\vldbavailabilityurl{https://github.com/lids-lab/steer}

\newcommand{\FtoP}{\ensuremath{\mathrm{F}{\to}\mathrm{P}}}
\newcommand{\PtoF}{\ensuremath{\mathrm{P}{\to}\mathrm{F}}}
\usepackage{array}
\newcolumntype{Y}{>{\centering\arraybackslash}X}
\usepackage{xcolor}
\usepackage{algorithm}
\usepackage[noend]{algpseudocode}          
\algrenewcommand\algorithmicindent{1em}    
\algrenewcommand\alglinenumber[1]{\scriptsize #1:}
\usepackage[capitalise]{cleveref}
\usepackage{multirow}
\usepackage{booktabs}
\usepackage[table]{xcolor}
\usepackage{placeins}
\usepackage{xltabular}
\usepackage{fvextra}
\usepackage{pgfplots}
\pgfplotsset{compat=1.18}

\usepackage{enumitem}
\setlist[itemize]{nosep,leftmargin=1.3em,labelsep=0.4em}
\setlist[enumerate]{nosep,leftmargin=1.3em,labelsep=0.4em}

\begin{document}
\title{STEER: Reducing Inference Cost in Relational Foundation Models through Semantically Informed Sampling}

\author{Abdalla Mohamed}
\affiliation{
  \institution{University of Texas at Arlington}
}
\email{axm0993@mavs.uta.edu}

\author{Ashraf Aboulnaga}
\affiliation{
  \institution{University of Texas at Arlington}
}
\email{ashraf.aboulnaga@uta.edu}

\begin{abstract}
Relational foundation models (RFMs) are pretrained once on a collection of relational databases and prediction tasks, and then applied zero-shot to previously unseen databases and tasks. To make a prediction for a target row, an RFM samples a neighborhood of rows linked to that row through foreign keys and uses this neighborhood as its inference context. Lowering inference cost is an important goal for any foundation model, and for RFMs this cost grows with the size of the context. The simplest ways to shrink the context 
is to drop some of the sampled rows, but this 
ignores the semantics of the database schema, so it is as likely to discard informative rows as uninformative ones. We propose STEER, a sampling approach that shrinks the inference context by concentrating it on the tables most relevant to the prediction task at hand. STEER obtains relevance information by prompting a large language model to rank the foreign-key edges of the database schema into relevance tiers for the given task, and then maps each tier to a probability of following that edge during traversal. Because the ranking uses only the schema, it is computed once per task and reused across all subsequent predictions, amortizing its cost. We evaluate STEER on three state-of-the-art RFMs (RT, RT-J, and Griffin) and show that it reduces inference context size by about 40\% on average while maintaining, and in some cases improving, prediction accuracy.

\vspace{-8pt}

\end{abstract}

\maketitle

\pagestyle{empty}
\thispagestyle{empty}

\ifdefempty{\vldbavailabilityurl}{}{
\begingroup\small\noindent\raggedright\textbf{Artifact 
Availability:}\\
The source code, data, and/or other artifacts have been made 
available at \url{\vldbavailabilityurl}.
\endgroup
}

\section{Introduction}
\label{sec:introduction}

\emph{Relational deep learning (RDL)} aims to train machine learning models directly on the structure of relational databases, without flattening the database into a single table and without the manual, dataset-specific feature engineering required by traditional tabular models~\cite{chen2016xgboost, shwartzziv2022tabular}. In RDL, a relational database is represented as a graph in which each row of each table is a node, and each foreign-key reference is an edge connecting two nodes~\cite{fey2024rdl}. 
A model can then propagate 
information along these edges, allowing the representation learned for 
a given row to be enriched with information from related rows. 
Early RDL models were shown to be effective across a wide range of classification and regression tasks, but they were trained for a specific database and task, requiring retraining from scratch for each new database or task~\cite{chen2025relgnn, dwivedi2026relgt}.

More recent work in RDL has focused on \emph{relational foundation models (RFMs)}, which are pretrained once on a collection of databases and can subsequently make zero-shot predictions for tasks on previously unseen databases, without updating their model weights. An early example is Griffin~\cite{wang2025griffin}, which is built around a graph neural network (GNN) architecture. Other examples include the closed-source, closed-weight, commercially available KumoRFM~\cite{fey2025kumorfm1} and KumoRFM-2~\cite{hudovernik2026kumorfm2} models. Another prominent example is the open-source Relational Transformer (RT)~\cite{ranjan2026rt}, which is built around a transformer architecture and has demonstrated strong performance across a range of evaluations. A recent follow-up, RT-J~\cite{ranjan2026rtj}, retains the RT architecture while introducing a large-scale pretraining recipe aimed at label-efficient prediction.

At inference time, an RFM must make predictions on databases it has never seen before and for which it has no prior knowledge of the structure or contents. Consequently, for each prediction task, the RFM constructs a \emph{sample} of the target database containing the information needed to make the prediction. The model sees the database only through the lens of this sample, so the sampler effectively determines what information is available to the model at inference time. This sample therefore serves as the \emph{context} for the prediction, and the cost of inference depends strongly on the size and complexity of this context.

Reducing inference cost is essential to the practical deployment of foundation models of all types, including RFMs. Unlike training, which is performed once, inference incurs a recurring cost for every prediction request, so even modest savings are multiplied across users, applications, and the lifetime of a deployed model. Accordingly, considerable attention in both academia and industry has focused on reducing the inference cost of large language models (LLMs), which are by far the dominant class of foundation models and the focus of most current AI activity~\cite{zhou2024survey}. Commercial LLM inference is typically priced by the number of input and output tokens processed, making context size not only a computational concern for the model provider but also a direct contributor to the monetary cost paid by model users. Approaches to reducing LLM inference cost include specialized inference hardware, such as Groq~\cite{abts2020thinkfast} and AWS Inferentia~\cite{awsinferentia}, as well as algorithmic techniques such as quantization~\cite{xiao2023smoothquant} and speculative decoding~\cite{leviathan2023speculative}. Another important line of work reduces inference cost by reducing the amount of context processed by the model, for example through context compression~\cite{li2023compressing} or selective attention~\cite{hooper2025squeezed}. In this paper, we pursue this latter direction for RFMs by constructing smaller contexts through semantically informed database sampling.

To motivate our sampling approach, we first describe how current RFM samplers construct inference context. An RFM prediction pertains to a particular row in the database, which we term the \emph{seed row}. Because RDL represents each row as a node in a graph, the seed row corresponds to a \emph{seed node} in the graph. RFM samplers construct the prediction context by sampling a neighborhood around this seed node. Starting from the seed node, the sampler traverses the graph along edges representing foreign-key links and adds the rows encountered during traversal to the sample. The process continues for a number of steps until a stopping condition is met, with the assumption that the resulting neighborhood provides informative context for prediction.

The cost of RFM inference grows with the size of the sampled context, since the computational cost of the model's forward pass increases with the amount of input it must process. Transformer-based RFMs consume the sample as a sequence of tokens, whereas GNN-based RFMs operate directly on the sampled subgraph. In either case, smaller samples reduce inference cost. A simple way to shrink the sample is to drop sampled rows at random. However, random dropping ignores the relevance of individual rows to the prediction task and can therefore substantially degrade prediction accuracy when informative rows are dropped.

Consider, for example, the Formula~1 database in the RelBench benchmark~\cite{robinson2024relbench}, the standard benchmark for relational deep learning. This database includes tables such as \texttt{drivers}, \texttt{results}, \texttt{standings}, and \texttt{qualifying}. Suppose the task is to predict whether a given driver (represented by a row in the \texttt{drivers} table) will fail to finish an upcoming race. The \texttt{results} table records each driver's previous race outcomes, including past instances of failing to finish, and therefore provides a more direct signal for this task than \texttt{standings} or \texttt{qualifying}, which describe a driver's ranking and grid position. A sampler that drops rows at random is just as likely to discard informative \texttt{results} rows as less useful \texttt{standings} or \texttt{qualifying} rows, potentially filling its limited context with information that contributes little to the prediction. A better approach, and the one we pursue in this paper, is to account for the expected usefulness of each table to the prediction task when deciding what information to sample.

Our approach is based on the observation that an important source of knowledge about the likely usefulness of tables is available without examining a single row of the database: the semantic structure of the schema itself. Table and column names are chosen to describe the real-world concepts they represent, and these descriptions can often indicate whether a table is likely to be relevant to a particular prediction task before any of its values are inspected.
Building on this observation, we propose \emph{STEER (Schema-guided, Token-Efficient Edge Ranking)}, a simple yet effective sampling approach that injects schema-derived semantic information about the usefulness of different tables for the prediction task into the sampling process. STEER replaces random row selection with a structured relevance ranking that governs the graph traversal itself. Rather than sampling broadly and then randomly dropping rows, STEER reduces sample size by actively choosing which edges to traverse and, consequently, which tables to sample. 

Another important observation underpinning STEER is that modern LLMs are a good source of deep semantic information about database schemas and how their table names, column names, and foreign-key connectivity relate to prediction tasks.  Therefore, STEER uses an LLM as a source of semantic knowledge and prompts this LLM to rank every edge leaving each table, in both traversal directions, into one of four tiers based on the relevance of this edge to the specific prediction task. Each tier is assigned a traversal probability, which the sampler uses to decide whether to follow an edge, rather than unconditionally traversing every edge, as in the default sampler. We also present an extension of STEER that identifies task-relevant columns, although the primary focus of STEER is on controlling graph traversal.

We implement STEER on the RT, RT-J, and Griffin RFMs, modifying each RFM's sampler, and we evaluate it on the RelBench benchmark, using tiered rankings generated by the Claude and GPT frontier LLMs. Across this evaluation, on average, STEER reduces sample size by roughly 40\% while matching, and sometimes slightly improving, the prediction accuracy of the baseline sampler operating with the full sample size. In contrast, random row dropping can (by construction) reduce sample size as much as needed, but this reduction comes at a steep cost in prediction accuracy.
In summary, our contributions are as follows:
\begin{itemize}
    \item A schema-aware LLM prompting strategy that produces tiered semantic relevance rankings for graph-traversal edges with respect to a given prediction task.
    
    \item A probabilistic sampling mechanism that uses these rankings to steer inference-time traversal, reducing sample size without sacrificing
    
    prediction accuracy.
    
    \item An optional column-level filtering mechanism that aims to remove columns that are unlikely to be informative for the prediction task.
    
    \item An empirical evaluation across three relational foundation models (RT, RT-J, and Griffin) and two frontier LLMs (Claude and GPT), showing that, on average, STEER reduces sample size by roughly 40\% while matching (or slightly improving) the prediction accuracy of baseline sampling with a full-size context.
\end{itemize}

\vspace{-8pt}
\section{Background: Sampling in RFMs}
\label{sec:background}

In this section, we provide a more detailed description of the sampling process used by RFMs. STEER modifies this process by incorporating semantic information into the sampling decisions.

Most predictive tasks of interest do not ask questions that a database already answers directly, but instead concern outcomes that have not yet occurred, such as whether a customer will churn next month or whether a Formula~1 driver will finish an upcoming race. To express such a prediction task to an RFM, a \emph{task table}~\cite{ranjan2026rt} is introduced alongside the existing database tables. Each row of the task table specifies
(1)~the entity for which a prediction is to be made (e.g., a customer or driver), specified through a foreign key pointing to that entity's row in the database;
(2)~a timestamp indicating the point in time at which the prediction is made; and
(3)~a label column containing the outcome to be predicted.
In a real deployment setting, the label value is not yet known and is therefore left blank. In an evaluation setting, the label is known but hidden from the model at prediction time and revealed only after the prediction is made, so it can be used for evaluation.

Different prediction tasks require different task tables. For example, predicting whether a customer will churn requires a different task table from predicting whether a Formula~1 driver will finish a race. A task table contains one or more rows corresponding to entities for which predictions are required, that is, rows whose label values are unknown or hidden from the model. It may also contain rows corresponding to \emph{example entities}, for which the prediction outcome is known and exposed to the RFM. These examples provide the model with information about the nature of the prediction task and play an important role in improving prediction accuracy.

The task table is linked to the database through the same graph representation used by RDL (every row is a node and every foreign-key reference is an edge). Because each task-table row contains a foreign key linking it to an entity in the database, the task-table row can itself be treated as another node in the graph, connected to the rest of the database through this foreign-key edge.

An alternative to explicitly constructing a task table is to present the prediction task through a declarative query language, such as the Predictive Query Language (PQL)~\cite{kocijan2026pql}.
PQL allows a user to express a prediction request as a query, much as SQL expresses a data-retrieval request. Such a query is ultimately compiled into the same underlying specification of an entity, timestamp, and prediction target that a task-table row encodes. We therefore describe RFM inference in terms of task tables throughout this paper.

To make a prediction, an RFM takes a task-table row as the starting seed row and  samples a portion of the surrounding database graph by traversing foreign-key links and adding encountered rows to the sample. Using the full graph for prediction is generally infeasible because the database graph may be too large to process in its entirety. The sampled portion of the graph is therefore provided to the model as input, and the model produces a prediction for the label associated with the seed row.

The traversal distinguishes two directions in which neighboring rows can be reached. First, it can move from a foreign-key \emph{child} row to the primary-key \emph{parent} row that it references. For example, the traversal may move from a \texttt{results} row to the corresponding \texttt{drivers} row. We refer to this as an \emph{\FtoP{}} link. Because a table has a fixed, typically small, number of foreign-key columns, and each foreign-key value references at most one parent row, \FtoP{} traversal has bounded fan-out.

Second, the traversal can move in the opposite direction, from a primary-key parent row to the foreign-key child rows that reference it. For example, it may move from a \texttt{drivers} row to all \texttt{results} rows that reference that driver. We refer to this as a \emph{\PtoF{}} link. This direction can have very high fan-out because a single parent row may be referenced by many child rows. Consequently, RFM samplers typically limit how many \PtoF{} neighbors are explored and select among the available candidates randomly. As a result, which particular rows ultimately appear in the sample is determined by chance rather than by their expected relevance to the prediction task. STEER modifies this by using semantic information to focus traversal on edges that are more likely to be useful.

The precise traversal strategy varies across RFMs, but all construct a bounded sample around the seed row and exclude \emph{temporally invalid} rows, that is, rows whose timestamps are later than the seed row's timestamp. This prevents \emph{temporal leakage}. Griffin~\cite{wang2025griffin}, which is GNN-based, samples a fixed number $B$ of neighbors per node for $K$ hops around the seed (typical values are $B=20$ and $K=2$). RT~\cite{ranjan2026rt}, which is transformer-based, constructs a context containing at most $L$ tokens using a modified breadth-first traversal over foreign-key links. Starting from the seed row, RT visits rows, adds their non-missing feature values as tokens to the context, and places neighboring rows on a \emph{frontier} for subsequent exploration. RT always includes \FtoP{} parent rows and subsamples \PtoF{} child rows subject to a width bound $w$, continuing until the context budget is exhausted or no additional rows remain reachable.
RT-J~\cite{ranjan2026rtj} retains RT's cell-token representation but replaces its basic neighborhood expansion with a retrieval-based sampling strategy. It performs multiple short random walks from the seed row, uses visitation frequency to identify promising rows, and then applies a local search to fill a bounded context window. STEER is designed to operate with any of these samplers because its decisions are made at the level of which foreign-key edge to follow while sampling.

\begin{figure}[t]
    \centering
    \includegraphics[width=\columnwidth]{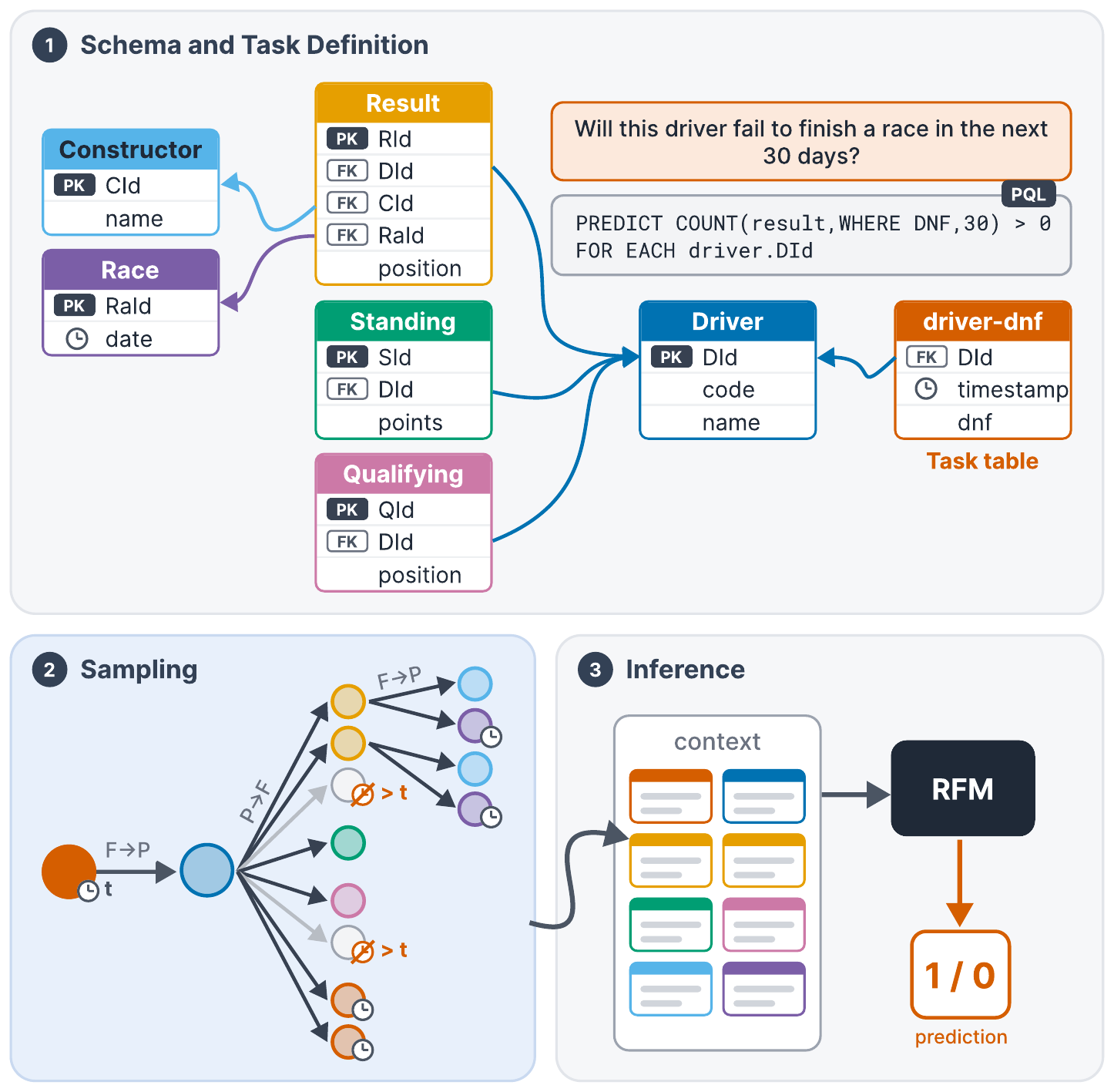}
    \caption{RFM inference on a relational database.
    \textbf{(1)}~The schema, with primary (PK) and foreign keys (FK), and
    a prediction task defined in English, in PQL, and as a task table whose
    foreign key references the entity being predicted for.
    \textbf{(2)}~Starting from the task row (seed), the sampler follows
    \FtoP{} and \PtoF{} links to collect related rows (node colors match
    the tables in~(1)); rows dated after the prediction timestamp~$t$ are
    excluded.
    \textbf{(3)}~The sampled rows form the context given to the RFM, which
    outputs the prediction.}
    \Description{A three-part figure. Part 1 shows a relational schema for
    Formula 1 with primary and foreign key links into a driver table, and
    the prediction task written in English, in PQL, and as a task table.
    Part 2 shows a graph of sampled rows expanding from the seed row, with
    rows dated after time t crossed out. Part 3 shows the sampled rows
    forming a context that is passed to a relational foundation model,
    which outputs a prediction of 1 or 0.}
    \vspace*{-18pt}
    \label{fig:rfm-overview}
\end{figure}

\Cref{fig:rfm-overview} illustrates the end-to-end RFM inference process on a small example database. The figure shows the database schema and task table, the portion of the graph sampled for a particular prediction, and the resulting model output.

\section{Controlling the Traversal}

\label{sec:control-points}

\begin{algorithm}[t]
\small
\caption{Adding STEER to an RFM sampler. \textsc{Select}, \textsc{Parents},
\textsc{Children}, and \textsc{Done} are abstract functions whose implementation defines the RFM's sampling steps. These functions are left unchanged by STEER. The \textcolor{blue}{blue} lines are STEER's only additions.}
\label{alg:steer}
\begin{algorithmic}[1]
\Require seed row $s$; row timestamps $\mathcal{T}$; \textcolor{blue}{probabilities $\Phi_{T,d}(e)\in[0,1]$ for prediction task $T$, direction $d$, and schema edge $e$}
\Ensure sample $S$
\State $S \gets \{\}$;\quad $F \gets \{s\}$ \Comment{sample, frontier}
\While{$F \neq \{\}$ \textbf{and not} \Call{Done}{$S$}}
    \State $r \gets \Call{Select}{F}$;\quad $F \gets F \setminus \{r\}$
    \State $S \gets S \cup \{r\}$
    \State $N_{\mathrm{FP}} \gets \Call{Parents}{r}$ \label{line:parents}
    \State \textcolor{blue}{$N_{\mathrm{FP}} \gets \Call{SteerFilter}{r, N_{\mathrm{FP}}, \FtoP}$} \Comment{\textcolor{blue}{filter parents}}
    \State $N_{\mathrm{PF}} \gets \{q \in \Call{Children}{r} \mid \mathcal{T}(q) \le \mathcal{T}(s)\}$ \Comment{no time leakage} \label{line:children}
    \State \textcolor{blue}{$N_{\mathrm{PF}} \gets \Call{SteerFilter}{r, N_{\mathrm{PF}}, \PtoF}$} \Comment{\textcolor{blue}{filter children}} 
    \State $F \gets F \cup N_{\mathrm{FP}} \cup N_{\mathrm{PF}}$ \Comment{add to frontier}
\EndWhile
\State \Return $S$
\Statex
\State \textcolor{blue}{\textbf{function} \textsc{SteerFilter}($r, N, d$)} 
\State \hspace{\algorithmicindent}\textcolor{blue}{\Return $\{\, q \in N \mid
    \mathrm{Uniform}(0,1) < \Phi_{T,d}(\mathrm{edge}(r,q)) \,\}$} 
\end{algorithmic}
\end{algorithm}

As discussed in \Cref{sec:background}, an RFM constructs its sample for inference by traversing the database graph outward from the seed row. STEER aims to control this graph traversal, and to steer it towards more useful and informative rows. We start by presenting the \emph{control points} at which STEER intervenes in the traversal.

The details of the traversal vary between RFMs, but all RFMs follow the overall abstract process described in \Cref{alg:steer}.
The blue lines in the algorithm are added by STEER and should be ignored for now. 
The steps \textsc{Select},
\textsc{Parents}, \textsc{Children}, and \textsc{Done} are abstract functions that are instantiated differently for different RFMs, and define the RFM's specific sampling algorithm.
The sampler maintains a frontier $F$ of rows to visit. At each step in the sampling loop, the sampler chooses one row $r$ from this frontier, determined by the \textsc{Select} function, and adds it to the sample $S$. Some or all the neighbors of row $r$ are visited, as determined by the \textsc{Parents} and \textsc{Children} functions, and added to the frontier $F$. This loop continues until the frontier is empty or some stopping condition, determined by the \textsc{Done} function, is met.

\Cref{alg:rt} shows an instantiation of \Cref{alg:steer} for RT~\cite{ranjan2026rt}. As before, ignore the blue lines for now. RT's sampler fills a context window $C$ with tokens sampled via graph traversal. It adds all the parents of a row $r$ and samples $w$ children. When faced with a choice among alternative rows, the sampler chooses at random, without regard to a row's relevance to the prediction task (lines~\ref{line:random1} and~\ref{line:random2}). This type of uninformed random sampling is performed by \emph{all} RFM samplers of which we are aware.

The place where STEER can intervene to fix this problem is the step in which the sampler decides which neighbors of a visited row are added to the frontier (lines~\ref{line:parents} and~\ref{line:children} of \Cref{alg:steer}).
STEER intervenes after these steps to guide the traversal towards rows that are more informative for the prediction task.
As described in \Cref{sec:background}, neighbors are reached through two types of links, \FtoP{} and \PtoF{}, which differ in their fan-out: an \FtoP{} traversal reaches at most one parent, whereas a \PtoF{} may reach many child rows.

Consider again the Formula~1 database from \Cref{sec:introduction} and the task of predicting whether a driver will fail to finish an upcoming race. The seed row in the task table reaches the corresponding row in the \texttt{driver} table through its foreign key. That \texttt{driver} row, in turn, connects through \PtoF{} links to several child tables, including \texttt{result}, \texttt{standing}, and \texttt{qualifying}. As discussed earlier, \texttt{result} rows are more directly informative for the task than \texttt{standing} or \texttt{qualifying} rows. STEER can guide the traversal to favor \PtoF{} links leading to \texttt{result} over those leading to the other two tables.

The same reasoning applies to \FtoP{} links. From a \texttt{result} row, for example, following the link to its parent \texttt{constructor} row may be informative because some constructors field more reliable cars than others. By contrast, following the link to the parent \texttt{race} row is likely less informative, since attributes such as race date or round have little direct bearing on mechanical failure. STEER can therefore favor the \FtoP{} link to \texttt{constructor} over the one to \texttt{race}.

STEER controls graph traversal by applying a \emph{traversal filter} after the sampler identifies the neighbor rows that it wants to add to the frontier. The filter has two variants, one for each traversal direction: an \emph{\FtoP{} edge filter} and a \emph{\PtoF{} edge filter}. Each filter is a function that is called after the sampler chooses its neighbors. This function decides, for every one of these neighbors, whether the edge from the current row $r$ to that neighbor should be traversed. This decision is made based on the estimated usefulness of the edge to the prediction task. If an edge is filtered out, its destination row is never added to the frontier and therefore never visited. Consequently, none of the neighbors reachable through that row are explored either, so filtering a single edge can prune the entire branch of the traversal rooted at that edge. The blue lines in Algorithms~\ref{alg:steer} and~\ref{alg:rt} show the definition of STEER's traversal filter and how it is added to the sampling loop of an RFM sampler. Adding STEER to the sampler of any other RFM will follow the same process.

\begin{algorithm}[t]
\small
\caption{Adding STEER to the RT sampler~\cite{ranjan2026rt}, an instance of
\Cref{alg:steer}. Note that the \textcolor{blue}{blue} lines added by STEER are unchanged from \Cref{alg:steer}. The way that STEER is added to a sampler does not change with the specifics of that sampler.
}
\label{alg:rt}
\begin{algorithmic}[1]
\Require seed row $s$;  row timestamps $\mathcal{T}$; context length $L$; width bound $w$; \textcolor{blue}{probabilities $\Phi_{T,d}(e)\in[0,1]$ for prediction task $T$, direction $d$, and schema edge $e$}
\Ensure set of cells $C$ (the context)
\State $C \gets \{\}$;\quad $F \gets \{s\}$;\quad $V \gets \{\}$ \Comment{context, frontier, visited}
\While{$F \neq \{\}$ \textbf{and} $|C| < L$} \Comment{\textsc{Done}: context length $L$ reached}
    \State $r \gets \Call{Select}{F}$;\quad $F \gets F \setminus \{r\}$
    \If{$r \in V$} \State \textbf{continue} \EndIf
    \State $V \gets V \cup \{r\}$;\quad $C \gets C \cup \mathcal{C}(r)$ \Comment{add $r$'s non-missing cells}
    \State $N_{\mathrm{FP}} \gets \mathcal{N}_{\FtoP}(r)$ \Comment{\textsc{Parents}: all parents of $r$}
    \State \textcolor{blue}{$N_{\mathrm{FP}} \gets \Call{SteerFilter}{r, N_{\mathrm{FP}}, \FtoP}$} \Comment{\textcolor{blue}{filter parents}} 
    \State $N_{\mathrm{PF}} \gets \{q \in \mathcal{N}_{\PtoF}(r) \mid \mathcal{T}(q) \le \mathcal{T}(s)\}$ \Comment{no time leakage}
    \State $N_{\mathrm{PF}} \gets \Call{RandomSample}{N_{\mathrm{PF}}, w}$ \Comment{\textsc{Children}: at most $w$} \label{line:random1}
    \State \textcolor{blue}{$N_{\mathrm{PF}} \gets \Call{SteerFilter}{r, N_{\mathrm{PF}}, \PtoF}$} \Comment{\textcolor{blue}{filter children}} 
    \State $F \gets F \cup N_{\mathrm{FP}} \cup N_{\mathrm{PF}}$ \Comment{add to frontier}
\EndWhile
\State \Return $C$
\Statex
\Function{Select}{$F$} \Comment{RT's \textsc{Select}}
    \State $R \gets \{r \in F \mid r \text{ was added via an F}{\to}\text{P link}\}$
    \If{$R = \{\}$}
        \State $R \gets \arg\min_{r \in F} \mathrm{HopDistance}(r, s)$ \Comment{rows closest to $s$}
    \EndIf
    \State \Return \Call{RandomSample}{$R$, $1$} \label{line:random2}
\EndFunction

\Statex
\State \textcolor{blue}{\textbf{function} \textsc{SteerFilter}($r, N, d$)} 
\State \hspace{\algorithmicindent}\textcolor{blue}{\Return $\{\, q \in N \mid
    \mathrm{Uniform}(0,1) < \Phi_{T,d}(\mathrm{edge}(r,q)) \,\}$} 

\end{algorithmic}
\end{algorithm}

We modified the samplers of three RFMs---RT, RT-J, and Griffin---to incorporate STEER's traversal filter. An important observation from this implementation effort is that all three models perform sampling within their machine-learning Python code, even though the data being sampled is relational and would ordinarily reside in a SQL DBMS. Adapting three distinct Python-based samplers required substantial model-specific implementation effort. As a direction for future work, we envision a unified RFM sampler implemented using SQL and exposed to different RFMs through a common API. Such an approach would simplify sampler implementation and make sampling strategies easier to modify and extend. It would also allow RFMs to operate directly on databases managed by a DBMS, without first exporting the data into a separate machine-learning pipeline, while benefiting from the query processing and optimization capabilities of the DBMS~\cite{khan2025sqlrdl}.

We now turn to the \textsc{SteerFilter} function. A key aspect of this function is that it makes the decision to retain or discard an edge in a \emph{probabilistic} manner. A function $\Phi_{T,d}(e)$ specifies, for prediction task $T$ and edge $e$ in direction $d$ (\FtoP{} or \PtoF{}), the probability that $e$ is relevant to $T$. \textsc{SteerFilter} flips a coin with this probability (i.e., conducts a Bernoulli trial) to decide whether to keep the edge $e$. Thus, edges assigned higher relevance are more likely to be retained, while less relevant edges are more likely to be pruned. Since $\Phi_{T,d}(e)$ is the probability that edge $e$ is included in the sample, we term it the \emph{inclusion probability}.

We made the design decision to drop edges in STEER probabilistically based on the following observation: \emph{edge relevance is rarely binary}. Although one could develop a hard classification of edges into either informative or uninformative for a given task, in practice edges are \emph{informative to varying degrees}. Even edges that are not directly relevant to the prediction target may still contribute useful signal. Returning to the Formula~1 example, \texttt{standing} and \texttt{qualifying} rows may provide some information for predicting whether a driver will fail to finish, even if they are less informative than \texttt{result} rows. Rather than eliminating all \texttt{standing} and \texttt{qualifying} rows, it is therefore preferable to reduce how often rows from these tables are sampled while preserving the more informative \texttt{result} rows. This finer-grained control is naturally achieved through probabilistic edge filtering.

The function $\Phi_{T,d}(e)$ is task-specific, so STEER must define two such relevance functions for every prediction task $T$, one for each of the \FtoP{} and \PtoF{} directions. It is clear that the effectiveness of STEER depends critically on defining good $\Phi_{T,d}(e)$ functions that accurately reflect the semantic relevance of an edge to a prediction task. We turn to this important topic next.
 
\begin{figure*}[t]
    \centering
    \includegraphics[width=\textwidth]{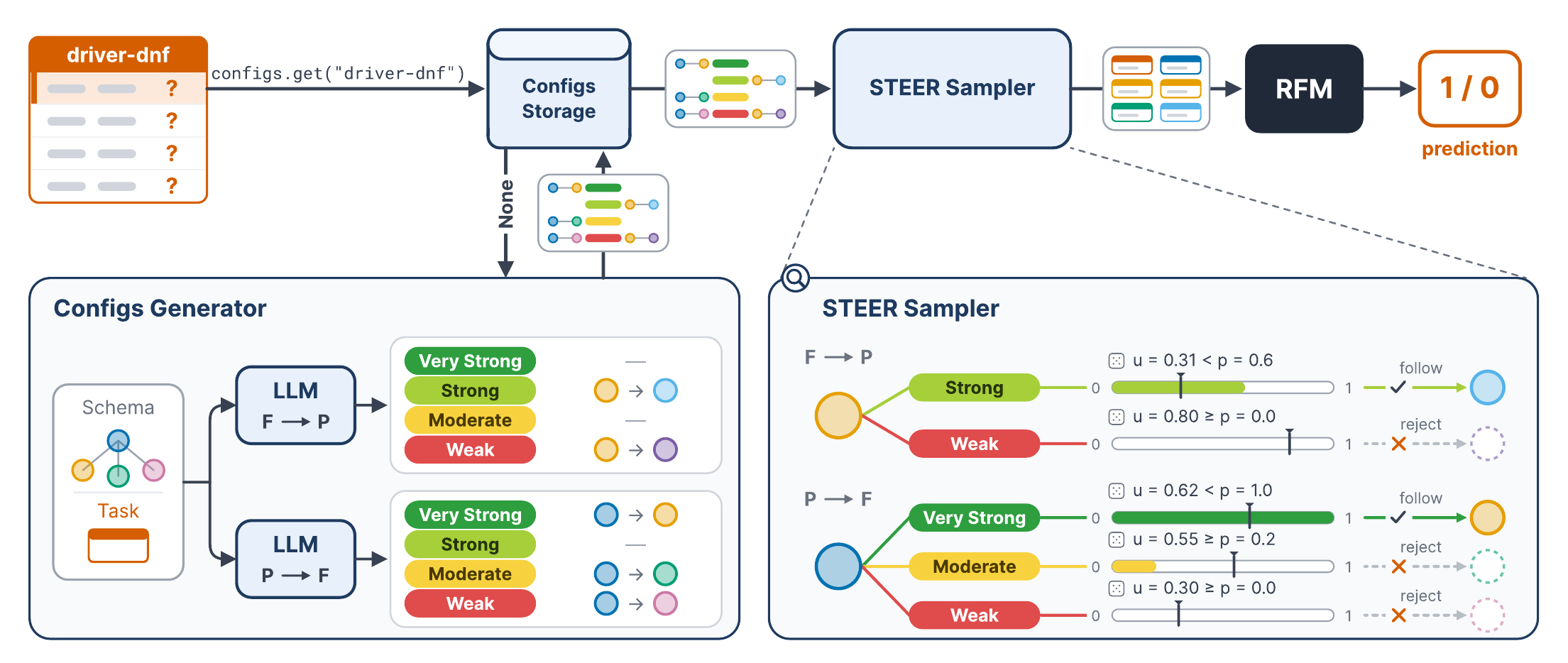}
    \caption{The STEER pipeline. A prediction request is a task
    table whose rows are prediction targets. STEER looks up the configuration
    of the task in the Configs Storage. If the storage returns \texttt{None},
    the Configs Generator generates a configuration with two LLM calls, one per traversal
    direction, that rank the schema edges into four relevance tiers. STEER saves this configuration in the Configs Storage, so that future invocations of this task will not result in LLM calls. The STEER Sampler then produces a sample using filtered traversal based on the configuration (\Cref{alg:steer}).
    The resulting sample is given to the RFM, which uses it for inference and outputs a prediction.}
    \Description{A pipeline diagram. On the left, a task table for the
    driver-dnf task has several rows, each marked with a question mark. An
    arrow labeled with a lookup of the task name leads to a Configs Storage. If
    the storage returns None, an arrow leads down to a Configs Generator, which
    contains two LLM calls, one for each traversal direction, that produce a
    tier for each schema edge, and the configuration returns to the storage. A
    configuration is passed to the STEER Sampler, which sends a sample to the
    RFM, which outputs a prediction of 1 or 0. An expanded view of the STEER
    Sampler shows, for a visited row, candidate edges with a number line from
    0 to 1 each: an edge is followed when the random number is below its
    probability and rejected otherwise.}
    \label{fig:pipeline}
\end{figure*}

\section{Determining Relevance}
\label{sec:determining-relevance}

\subsection{Estimating Semantic Relevance with LLMs}
\label{sec:semantic-relevance}

In this section, we discuss how to decide if an edge is relevant to a given prediction task. In \Cref{sec:steering-sampler}, we turn this decision into a $\Phi_{T,d}(e)$ function.
Deciding whether an edge is worth traversing
is fundamentally a question of meaning: does the table
at the other end of
this edge describe something related to the prediction target? This kind of
question can often be answered without inspecting a single row of the
database. Table and column names are typically chosen to describe the real-world
concepts they represent, and that description alone is often enough to judge
whether a concept is likely to matter for a given prediction.
Semantic relevance of this kind has been 
used in other settings such as training task-specific RDL models~\cite{ding2026gelgt}.

LLMs are well suited to producing this kind of semantic judgment. They have
been shown to interpret schema-level information effectively in other data
management tasks that similarly require judging correspondences or relevance
from names and descriptions alone, such as schema matching~\cite{parciak2024schemamatching}
and text-to-SQL~\cite{li2024codes}.
LLMs have also been used to directly
guide sampling for inference in GNNs~\cite{afandi2026kgwise}. We use an LLM for this purpose in STEER, through a purpose-built prompting procedure that elicits a structured relevance judgment for every traversal edge in the schema.

The LLM in STEER is given the full database schema, and it judges \emph{schema edges}. Each foreign-key column defines one
schema edge $e$, which the sampler can traverse in two directions
$d \in \{\FtoP, \PtoF\}$.
A table with several foreign keys to the same
parent defines several edges, and a foreign key that references its own
table defines a self-referential edge. We rank edges rather than tables
because the value of reaching a table depends on where the sampler comes
from: two edges that lead to the same table can differ in relevance
depending on the table from which each is reached.

\vspace{-8pt}
\subsection{Prompting Strategy}
\label{sec:prompting-strategy}

A natural way to obtain a judgment about the relevance of an edge is to use a simple prompt: give the LLM
the database schema and the prediction task, ask it directly which edges
are relevant, and accept a binary include-or-exclude answer for each
candidate. 
Our experiments show that modern LLMs give reasonable results even with this simple prompt (\Cref{sec:prompting-ablation}).
However, there is room for improvement, and the prompt used by STEER builds on this simple prompt in two ways.

First, it asks the LLM to reason in explicit steps, first about the
database schema, then about the prediction task, and only then about how
each edge relates to them, rather than answering directly from the raw
schema and task description. Judging the relevance of an edge depends on
understanding the schema as a whole and the task, so we expect this
ordering to help. Guiding a language model through intermediate reasoning
steps has been shown 
to improve the reliability of its
outputs on tasks that require combining several pieces of information before
reaching a conclusion~\cite{wei2022cot}.

Second, instead of a binary decision, the prompt asks the LLM to assign
each edge to one of four ordered \emph{tiers} of relevance: \emph{Very Strong},
\emph{Strong}, \emph{Moderate}, or \emph{Weak}. A binary judgment
cannot distinguish an edge that is worth following occasionally from
one that is not worth following at all from one that is worth following always. A tiered ranking can make such a fine-grained distinction.
Comparative, graded judgments of this kind have been found to
yield more consistent outputs from LLMs
than isolated binary
ones~\cite{qin2024pairwise}.

The STEER prompt is organized into four components: (1)~a task objective
instructing the LLM to rank edges by their semantic relevance to the
prediction target, (2)~the reasoning steps described above, (3)~a set of rules
constraining the output to valid schema elements, to prevent the LLM from
hallucinating tables that do not exist, and (4)~the concrete database and task
definition, including table names, column names, primary and foreign keys, and the target
column. 
The LLM is consulted once for each new database and prediction task.
Two prompts are issued, one for each traversal direction, and the resulting configuration is reused
across every subsequent prediction on that task, so the cost of querying the
LLM is incurred once per task rather than once per prediction. The simple and STEER prompt templates are listed in Appendix~\ref{sec:appendix-prompts}.
\vspace{-8pt}
\section{Inclusion Probabilities}
\label{sec:steering-sampler}

The LLM produces a configuration that assigns a relevance tier to every schema edge in each traversal direction. What remains is converting these tiers into inclusion probabilities $\Phi_{T,d}(e)$ for the traversal filters (\Cref{sec:control-points}). We adopt the following simple approach for this conversion: for every task, we assign an inclusion probability to each relevance tier in a given traversal direction. Since we have four relevance tiers and two traversal directions, this is a total of eight probabilities for each task. All edges in a given tier get assigned that tier's inclusion probability.

Thus, we define two functions: (1)~$\mathrm{tier}_{T,d}(e)$, which is the relevance tier that the LLM assigns to edge $e$ in direction $d$ for task $T$, and (2)~$\pi(v)$ which maps relevance tier $v$ to a probability. Using these two functions, we define
$\Phi_{T,d}(e)= \pi\big(\mathrm{tier}_{T,d}(e)\big)$.

We also refer to the tier-to-probability mapping function $\pi(v)$ as the \emph{probability profile} of a task. It is possible to tune the probability profile for every task, but this is costly since it requires performing a search over the possible values of the mapping function, and running the full RFM inference pipeline for each point in this search. Instead, our default approach in STEER is to use a \emph{fixed} probability mapping for all tasks and navigation directions. Specifically, \emph{Very Strong} maps to $1.0$, \emph{Strong} to $0.6$, \emph{Moderate} to $0.2$, and \emph{Weak} to $0.0$. We show in \Cref{sec:prob-tuning} that this choice yields good results, and probability tuning is warranted in only limited cases.

We can now present another reason for adopting probabilistic edge filtering, in addition to the edge informativeness reason discussed in \Cref{sec:control-points}: probabilistic filtering makes STEER more robust to errors made by the LLM. A binary approach, where all edges in high relevance tiers are sampled and all edges in low relevance tiers are dropped, increases the impact of LLM errors. An edge placed one tier too low can be permanently cut off from the sample, however useful it might actually have been, while an edge placed one tier too high would permanently take up room in the context that it does not deserve. The probabilistic approach mitigates this effect and turns the LLM's ranking from a fixed verdict into a graded influence on traversal: the higher the tier of an edge, the larger the share of its candidates that the sampler follows, and an edge assigned to a neighboring tier by mistake is followed somewhat more or less often, rather than being included or excluded outright.

\vspace{-8pt}
\section{STEER Pipeline}
\label{sec:pipeline}

This section puts the components of STEER together into an end-to-end pipeline, shown in \Cref{fig:pipeline}. The pipeline starts when an RFM using STEER is presented with a prediction request in the form of a task table whose rows are prediction targets. The pipeline has two parts that run at different frequencies: obtaining the configuration using the LLM, which happens once per database and task, and sampling with the configuration, which happens for every row of the task every time the task is invoked.

STEER conserves calls to the LLM by obtaining a configuration for each task and database once, and reusing the configuration on subsequent invocations of the same task. To accomplish this, STEER stores all computed configurations in a \emph{Configs Storage}, which is a key-value store where the key is a task and the value is the STEER configuration for that task. For each prediction request, STEER first performs a lookup in the Configs Storage using the task table as the key (for example, \texttt{configs.get("driver-dnf")}). If a configuration is found for the task, STEER passes it directly to the sampler. If the storage returns \texttt{None}, this indicates the first time a task is seen, and the \emph{Configs Generator} creates the configuration. It issues two LLM calls, one for each traversal direction (\FtoP{} and \PtoF{}), each with the schema and the task definition (Section~\ref{sec:prompting-strategy}), and creates a configuration consisting of a relevance tier for every schema edge. The configuration is saved in the Configs Storage before it is passed on to the sampler. A configuration depends only on the schema and does not change for a given task, regardless of the rows being predicted. The STEER sampler (\Cref{alg:steer}) then uses the configuration and probability profile to build a sample for each row in the task table, and passes this sample to the RFM which produces a prediction. The cost of STEER is, therefore, a lookup in the Configs Storage for a previously seen task, and two LLM calls when a task is seen for the first time.

\vspace{-12pt}
\section{Extending STEER: Column Filter}
\label{sec:extensions}

The mechanism that STEER uses to filter 
edges applies equally well to a second
point in sampling: once a row has been reached, not every one of its columns
is equally useful to include in the sample. A \texttt{result} row, for
instance, contains a column recording the outcome of a driver's race,
including whether they failed to finish, which is very relevant to a
did-not-finish prediction. It also contains columns such as a driver's starting
grid position or lap time, which are useful but comparatively less informative for this
particular task. The row also contains columns such as the result identifier or a comment that carry no
predictive signal at all. Writing every column of a visited row to the sample without
distinction spends context budget on cells that do little to help the prediction.
Unlike the edge filters, which reduce the sample by changing the rows that are
visited, the column filter changes what is written to the context once a row has already
been visited: withholding a column removes the context cells that column would have
contributed, which shrinks the context and leaves room for more
informative columns, or for more rows, within the same context length.

The column filter applies only to transformer-based RFMs such as RT and
RT-J, in which each cell of a row is a separate element of the context, so a
single column can be individually withheld. In graph-based RFMs such as
Griffin, a row is represented as a single node, so there is no per-column
unit to remove.

We address columns the same way we address traversal. An LLM ranks every
column of every table into one of the same four tiers, using the prompting
strategy described in \Cref{sec:prompting-strategy}, and the tier of
a column $c$ is mapped to an inclusion probability
$\Phi_{T,d}(c)= \pi\big(\mathrm{tier}_{T,d}(c)\big)$, using the same mapping~$\pi$
as for edges. When a row is visited, each of its columns is
written to the context with the probability defined by $\Phi_{T,d}(c)$, and the writing decision for each column is decided
independently by a coin flip (Bernoulli trial).

We identify a relatively common case where the column filter can fall short: the LLM derives its ranking entirely from the column name and does not inspect the values that the column actually stores, so it may recommend a column where a large fraction of the values are actually nulls. Such a column would consume context budget while offering little practical signal, a fact that is not knowable from the column name.
This phenomenon reflects a more general pattern in how LLM judgments are
best used: they are most reliable when used to guide a downstream
procedure grounded in the actual data rather than being trusted
directly~\cite{marchesin2026stratification}.

We correct for this with a lightweight statistical pass over the database:
for every column $c$, we compute its \emph{null rate} from the data, defined as the fraction of rows in the table for which this column is null.
If the null rate of a column exceeds a fixed
threshold, we demote it to the \emph{Moderate} tier, regardless of the
tier the LLM originally assigned it. We settle on \emph{Moderate}
because a column that is mostly null may still carry some
signal in the rows where a value happens to be present, so its relevance is
better described as uncertain rather than clearly high or clearly low.
Like other aspects of the relevance configuration, this correction is computed once per new task and subsequently reused.

The column filter is independent of the edge filters and can be used with or
without them.
Since the column filter works only for transformer-based RFMs and not graph-based RFMs, and since STEER focuses primarily on the navigation during sampling, we consider the column filter to be an \emph{optional extension} to STEER. We show in \Cref{sec:column-filter-ablation} that this extension can be useful, but it is not as robust as the traversal element of STEER.

\section{Experiments}
\label{sec:experiments}

\subsection{Experimental Setup}
\label{sec:experimental-setup}

\noindent\textbf{Hardware.}
All experiments were conducted on a GPU server with 8 NVIDIA H100 GPUs with 80~GB of memory each, 2 AMD Genoa CPUs, and 2.3~TB of RAM.

\vspace{4pt}\noindent\textbf{Benchmark.} We evaluate on the RelBench benchmark~\cite{robinson2024relbench},
using every database from RelBench v1 except \texttt{rel-event}, which
we exclude due to reported temporal leakage issues~\cite{ranjan2026rt}, together with two
databases added in RelBench v2~\cite{gu2026relbenchv2},
\texttt{rel-arxiv} and \texttt{rel-ratebeer}. This gives a total of 24
tasks across 8 databases, spanning both classification and regression.

\vspace{4pt}\noindent\textbf{RFM Details.} We evaluate STEER on three relational foundation
models: RT~\cite{ranjan2026rt}, RT-J~\cite{ranjan2026rtj}, and Griffin~\cite{wang2025griffin}. We evaluate on RT since the best RFM accuracy at the time of writing is achieved by models from the RT family. RT-J is the most recent model from this family, and the most accurate in recent evaluations.
Griffin is an older, less accurate model, but we include it to demonstrate that STEER works with any RFM, including GNN-based models like Griffin, not just transformer-based models like RT or RT-J.

For RT and RT-J, we used the code and model checkpoints provided by the authors.\footnote{\url{https://github.com/stanford-star/relational-transformer}}
The provided RT checkpoints do not include the new \texttt{rel-arxiv} and \texttt{rel-ratebeer} databases, so we trained RT from scratch using the code provided by the authors on a corpus that includes these two databases in addition to the six databases on which RT was originally trained. We use the \emph{leave-one-DB-out} curriculum from the RT paper~\cite{ranjan2026rt}: when conducting inference on a database, we use a model checkpoint trained on the remaining 7 databases, so the inference database is not seen by the model during training and the predictions are zero-shot.
RT-J is trained on a very large corpus, and we use the model checkpoint provided by the authors.
For Griffin, we use a model checkpoint provided by the RT authors via personal communication, since they modified the model to enable zero-shot predictions.

We modified the samplers of all three RFMs to use STEER during inference. RT and RT-J are both configured to sample with a context length of $1024$ tokens. Griffin is configured to sample with a fanout of $20$ and a traversal depth of $2$ hops. All our code and model checkpoints are available.\footnote{\url{https://github.com/lids-lab/steer}}

\vspace{4pt}\noindent\textbf{LLMs.} We use two frontier LLMs to produce STEER's tiered ranking: Anthropic's Claude Sonnet 5 and OpenAI's GPT o3. Unless otherwise noted, the LLMs are queried through their APIs using the smart prompting strategy described in Section~\ref{sec:prompting-strategy}, at high reasoning effort. STEER requires two LLM API calls the first time a task is seen, one for 
$\FtoP$ and one for $\PtoF$.

\vspace{4pt}\noindent\textbf{Accuracy Metrics.} Classification tasks are evaluated with
AUROC and regression tasks with $R^2$. Both are reported as
percentages (i.e., multiplied by $100$), with higher values indicating
better accuracy for both metrics. Every reported result, including
sample size, is an average over an evaluation budget of $1280$ items per
task.
For all models, one inference run on this number of items typically takes less than 5 minutes on 1 GPU on our server.

\begin{figure}[t]
    \centering
    \includegraphics[width=\linewidth]{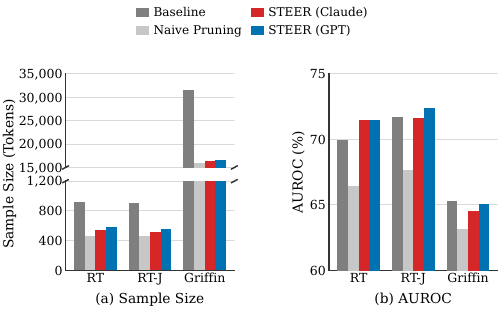}
    \caption{Classification results: average sample size and test accuracy (AUROC).}
    \Description{Two side-by-side bar charts comparing Baseline, Naive Pruning, STEER with Claude, and STEER with GPT across RT, RT-J, and Griffin. The left panel shows average sample size in tokens and the right panel shows average AUROC.}
    \vspace*{-8pt}
    \label{fig:main-cls}
\end{figure}

\vspace{4pt}\noindent\textbf{Sampler Configurations and Sample Size.} We compare four sampler configurations: \emph{Baseline} 
constructs a sample through unmodified random sampling at each model's 
default budget. \emph{Naive pruning} constructs a sample the same 
way, then discards half of the resulting rows uniformly at random, 
isolating the effect of simply shrinking the sample from the effect of 
carefully choosing rows to keep. \emph{STEER (Claude)} and \emph{STEER 
(GPT)} apply STEER's traversal filter, using tiered rankings from Claude and GPT, respectively.

For RT and RT-J, we report sample size as the
number of tokens placed in the context window by the sampler.
Griffin does not
tokenize its input the same way, so we instead report an estimated
token count, obtained by converting Griffin's sampled node count using the
average number of columns per row across the database.

\vspace{4pt}\noindent\textbf{Default Probability Profile.} Unless otherwise noted, STEER
uses a single, universal tier-to-probability mapping applied
identically across every task, model, and navigation direction: \emph{Very
Strong} maps to $1.0$, \emph{Strong} to $0.6$, \emph{Moderate} to
$0.2$, and \emph{Weak} to $0.0$.

\vspace{4pt}\noindent\textbf{Detailed Results.} This section reports the summary of the experimental results. The full task-by-task results are reported in Appendix~\ref{sec:appendix-full-results}.

\subsection{Main Result: Effectiveness of STEER}
\label{sec:main-results}

\begin{figure}[t]
    \centering
    \includegraphics[width=\linewidth]{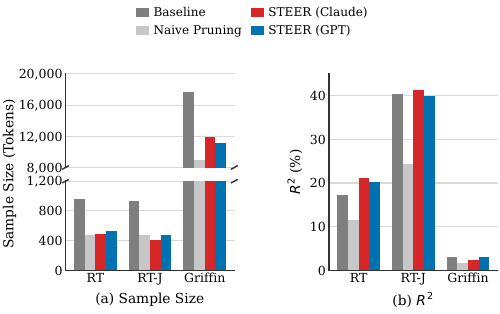}
    \caption{Regression results: average sample size and test accuracy ($R^2$).}
    \Description{Two side-by-side bar charts comparing Baseline, Naive Pruning, STEER with Claude, and STEER with GPT across RT, RT-J, and Griffin. The left panel shows average sample size in tokens and the right panel shows average R-squared.}
    \vspace*{-12pt}
    \label{fig:main-reg}
\end{figure}

We compare the four sampler configurations described above on all 24 tasks. 
Figures~\ref{fig:main-cls} and~\ref{fig:main-reg}
report the average sample size for each
configuration and the corresponding average test accuracy on classification tasks and regression tasks, respectively.

Let us examine the sampler configurations one by one.
The baseline sampler for RT and RT-J does not, on average, fill the entire $1024$-token context window, indicating that it often exhausts the available eligible tokens before reaching the context limit. 
The resulting sample sizes are similar across the two models and across classification and regression tasks, demonstrating the consistency of the sampling procedure. Under baseline sampling, RT-J achieves slightly higher accuracy than RT on classification tasks and much higher accuracy on regression tasks, consistent with RT-J being a larger and more capable model. Griffin, in contrast, uses samples that are roughly an order of magnitude larger than those of RT and RT-J while achieving considerably lower accuracy. This pattern persists across all sampler configurations and is consistent with prior comparisons showing that Griffin underperforms RT~\cite{ranjan2026rt}. As noted earlier, we include Griffin to demonstrate that STEER applies to GNN-based RFMs as well as transformer-based ones.

Naive pruning reduces sample size by exactly 50\% in all cases, by construction, since it always discards half the sample. However, this reduction comes at a substantial cost in prediction accuracy. Naive pruning performs markedly worse than all other sampler configurations. For RT-J, for example, naive pruning reduces classification AUROC by $4.10$ percentage points relative to the baseline sampler and reduces regression $R^2$ by $16.09$ percentage points, representing a severe degradation in prediction quality.

STEER also substantially reduces sample size relative to the baseline sampler across all three RFMs and with both LLMs. For RT and RT-J, sample size is reduced by $36.9$--$43.3\%$ on classification tasks and by $32.8$--$56.7\%$ on regression tasks. For Griffin, the corresponding reductions are $47.5$--$48.2\%$ for classification and $32.8$--$37.0\%$ for regression. These reductions are comparable to the $50\%$ reduction achieved by naive pruning. In one case, RT-J on regression tasks using Claude, STEER reduces sample size by $56.7\%$, exceeding the reduction produced by naive pruning.

Crucially, STEER achieves these reductions with little or no loss in prediction accuracy. At worst, STEER incurs only a small decrease relative to the baseline sampler, and in several cases it improves accuracy. For RT, classification AUROC improves by as much as $1.48$ percentage points and regression $R^2$ by as much as $3.89$ percentage points. RT-J likewise exhibits accuracy improvements under STEER. We attribute these gains to STEER's ability to remove less useful information from the sample, yielding a more focused and task-relevant inference context. Griffin exhibits only minor accuracy degradation across both task types and both LLM sources, with a maximum decrease of $0.78$ percentage points. Its performance therefore remains close to the baseline and substantially better than naive pruning on the same model.

In summary, the central empirical finding of this paper is that \textbf{STEER substantially reduces sample size without sacrificing prediction accuracy}. In contrast, naive pruning reduces sample size only at the cost of significant accuracy degradation. These results demonstrate that task relevance is critical when reducing the inference context, and that STEER's LLM-based semantic ranking effectively captures this relevance. We also find that both Claude and GPT are effective sources of semantic information for STEER. Finally, comparing RFMs, we find that RT and RT-J consistently achieve strong prediction accuracy with relatively compact samples, whereas Griffin requires substantially larger samples while achieving lower accuracy.

We next present ablation experiments examining the individual components of STEER. We ran these experiments with all three RFMs and both LLMs and observed the same qualitative patterns throughout. However, because Griffin is substantially weaker than RT and RT-J in our evaluation, and because Claude and GPT produce similar results, we report the remaining experiments only for the transformer-based RT and RT-J models and use Claude as the LLM.

\subsection{Ablation: Direction of Traversal}
\label{sec:direction-ablation}

\begin{figure}[t]
    \centering
    \includegraphics[width=\linewidth]{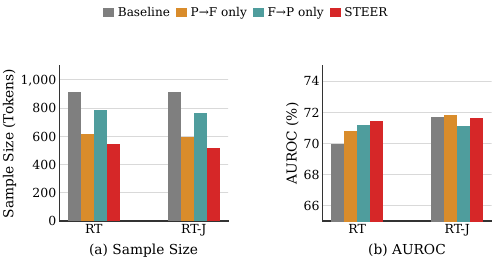}
    \caption{Classification results for direction of traversal ablation: average sample
    size and test accuracy (AUROC).}
    \Description{Two side-by-side bar charts comparing Baseline, P-to-F only, F-to-P only, and STEER across RT and RT-J. The left panel shows average sample size in tokens and the right panel shows average AUROC.}
    \vspace{-8pt}
    \label{fig:direction-cls}
\end{figure}

\begin{figure}[t]
    \centering
    \includegraphics[width=\linewidth]{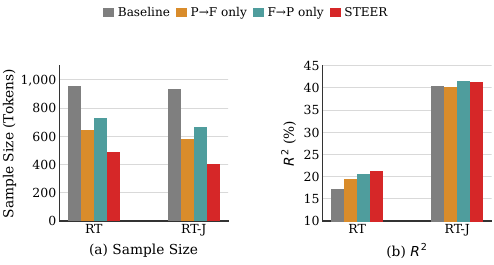}
    \caption{Regression results for direction of traversal ablation: average sample
    size and test accuracy ($R^2$).}
    \Description{Two side-by-side bar charts comparing Baseline, P-to-F only, F-to-P only, and STEER across RT and RT-J. The left panel shows average sample size in tokens and the right panel shows average R-squared.}
    \vspace{-8pt}
    \label{fig:direction-reg}
\end{figure}

\begin{table*}[t]
\caption{Prompting strategy ablation: average test accuracy for classification and regression, comparing simple prompting with STEER's smart prompting. Results are shown for Claude and GPT with high and default reasoning effort. Within each simple/STEER pair, bold indicates the higher accuracy.}
\vspace*{-4pt}
\label{tab:prompting-summary}
\centering
\small
\setlength{\tabcolsep}{4pt}
\begin{tabular}{l c|cc|cc|cc|cc}
\toprule
\multirow{2}{*}{RFM} & \multirow{2}{*}{Base.} & \multicolumn{2}{c|}{Claude High} & \multicolumn{2}{c|}{Claude Default} & \multicolumn{2}{c|}{GPT High} & \multicolumn{2}{c}{GPT Default} \\
& & Simple & STEER & Simple & STEER & Simple & STEER & Simple & STEER \\
\midrule
\multicolumn{10}{l}{\textbf{Classification (AUROC, \%)}} \\
RT    & 69.98 & 71.21 & \textbf{71.46} & 70.58 & \textbf{71.30} & 69.64 & \textbf{71.44} & 70.04 & \textbf{71.27} \\
RT-J  & 71.72 & 70.83 & \textbf{71.64} & 71.34 & \textbf{71.52} & 70.60 & \textbf{72.37} & 70.77 & \textbf{71.92} \\
\midrule
\multicolumn{10}{l}{\textbf{Regression ($R^2$, \%)}} \\
RT    & 17.33 & 17.59 & \textbf{21.21} & 18.05 & \textbf{21.51} & 16.35 & \textbf{20.31} & 17.14 & \textbf{20.60} \\
RT-J  & 40.43 & 34.12 & \textbf{41.22} & 34.22 & \textbf{41.03} & 33.85 & \textbf{39.91} & 35.61 & \textbf{39.28} \\
\bottomrule
\end{tabular}
\end{table*}

STEER's traversal filter operates on both traversal directions simultaneously: \textit{\FtoP{}} links, which move from a child row to a parent row that it references, and \textit{\PtoF{}} links, which move from a parent row to child rows that reference it (Section~\ref{sec:background}). This experiment isolates the contribution of each direction by applying STEER's tiered ranking to only one direction at a time while leaving the other unfiltered. The goal is to determine whether one direction contributes more than the other and whether filtering both directions together is more effective than filtering either direction alone.

Figures~\ref{fig:direction-cls} and~\ref{fig:direction-reg}
report the average sample size and corresponding test accuracy for each
traversal direction alone and for STEER, which combines the two directions. The figures show results for classification and regression tasks,
respectively.

Filtering either direction alone reduces sample size. \PtoF{} filtering reduces the number of tokens by $32.2$--$35.0\%$, while \FtoP{} filtering yields reductions of $14.5$--$29.3\%$ across both models and task types. Neither direction alone, however, matches the reduction achieved by filtering both directions together ($40.3$--$56.7\%$). The two directions are therefore complementary: each targets a different part of the redundancy in a 
random sample, and combining them removes more than either removes on 
its own.

Prediction accuracy exhibits a similarly favorable pattern. For RT, filtering \PtoF{} links improves accuracy on both classification and regression tasks, while filtering \FtoP{} links produces slightly larger improvements. STEER, which filters both directions, yields the largest accuracy gains. For RT-J, all three configurations remain close to baseline accuracy, with only small increases or decreases.

Overall, both traversal directions contribute meaningfully to reducing sample size, and neither direction results in a loss of accuracy. Using both directions together yields the largest reduction in sample size and the highest prediction accuracy.

\begin{table}[t]
\caption{Column filter ablation: average test accuracy and sample 
size for classification and regression, comparing baseline sampling, STEER without column filter, and STEER with column filter. Bold indicates the highest test score in a row and the highlighted cell indicates the lowest token count.}
\label{tab:colfilter-summary}
\centering
\small
\setlength{\tabcolsep}{6pt}
\begin{tabular}{l cc cc cc}
\toprule
\multirow{2}{*}{RFM} & \multicolumn{2}{c}{Baseline} & \multicolumn{2}{c}{STEER} & \multicolumn{2}{c}{STEER + Col.\ Filter} \\
& Test $\uparrow$ & Tok. $\downarrow$ & Test $\uparrow$ & Tok. $\downarrow$ & Test $\uparrow$ & Tok. $\downarrow$\\
\midrule
\multicolumn{7}{l}{\textbf{Classification (AUROC, \%)}} \\
RT   & 69.98 & 915.6 & \textbf{71.46} & 546.6 & 70.82 & \cellcolor{green!15}467.4 \\
RT-J & 71.72 & 912.6 & 71.64 & 517.7 & \textbf{72.20} & \cellcolor{green!15}450.6 \\
\midrule
\multicolumn{7}{l}{\textbf{Regression ($R^2$, \%)}} \\
RT   & 17.33 & 954.5 & 21.21 & 489.1 & \textbf{26.60} & \cellcolor{green!15}409.2 \\
RT-J & 40.43 & 937.2 & \textbf{41.22} & 405.8 & 38.11 & \cellcolor{green!15}353.9 \\
\bottomrule
\end{tabular}
\end{table}

\subsection{Ablation: Prompting Strategy}
\label{sec:prompting-ablation}

STEER's smart prompting strategy (Section~\ref{sec:prompting-strategy}) 
combines structured reasoning steps with a tiered output format. This 
experiment compares it against simple prompting, which asks for the 
same relevance judgment directly with no reasoning steps and a binary 
output. Our goal is to compare the two prompting strategies across multiple LLM variants. Therefore, we conduct this experiment with both Claude and GPT. To investigate whether a powerful LLM will be substantially better than a simpler LLM, we use two settings for the reasoning effort in both models: 
\emph{High}, the setting used throughout the rest of this paper, and 
\emph{Default}, the LLM's standard setting. Table~\ref{tab:prompting-summary} 
reports average test accuracy.

The main takeaway from the table is that STEER's smart prompting outperforms simple prompting in every setting, thus justifying STEER's use of smart prompting. The gap is modest on classification, typically under one percentage point, indicating that smart prompting has low benefit for these tasks.
However, the gap is substantial on regression: RT-J's $R^2$ under simple 
prompting falls around $6$ percentage points below the baseline across all LLMs,
while STEER stays near or above the baseline, thereby providing strong support for using smart prompting.

A second takeaway from the table is that STEER works well with all LLMs. Claude is slightly more accurate than GPT for regression, but both LLMs are sufficiently accurate. There is no big difference between high and default reasoning effort, indicating that STEER is effective even with less powerful, less expensive LLMs.

\subsection{Ablation: Column Filter}
\label{sec:column-filter-ablation}

Section~\ref{sec:extensions} proposes extending STEER's tiered 
ranking from edges to individual columns, filtering which columns of a 
visited row are written into the sample. This experiment evaluates the 
column filter applied on top of STEER's traversal filter. (In this experiment and the next, we return to using only Claude-based rankings.) 
Table~\ref{tab:colfilter-summary} reports 
average test accuracy and sample size for baseline sampling, STEER without the column filter, and STEER with the column filter. The table reports results for both classification and regression.

As expected, adding the column filter on top of 
STEER's traversal filter reduces sample size in every case, by 
an additional $5.6$--$9.8$ percentage points compared to traversal filtering 
alone. 
The effect on prediction accuracy is not as uniformly positive across models. On RT, the column filter improves $R^2$ 
substantially ($5.39$ percentage points compared to STEER without the column filter) and leaves classification accuracy close to STEER without the column filter. On RT-J, the column filter improves classification slightly ($+0.56$ percentage points compared to STEER without the column filter) but drops regression accuracy by $3.11$ percentage points.

Overall, the column filter has a directionally positive effect: it improves the sample size in all cases and substantially improves prediction accuracy in some cases. However, there are cases where it hurts prediction accuracy. 
Therefore, we leave the column filter as a promising extension to STEER that is useful but risky, and thus warrants further investigation as part of future work.

\begin{table*}[t]
\centering
\small
\caption{Probability tuning ablation: classification tasks (AUROC~\%, higher is better). For each task and model, we report test accuracy and number of tokens with baseline sampling, STEER with the default probability profile, and STEER with a probability profile tuned for the specific task. Within each row, bold indicates the highest test score and the highlighted cell indicates the lowest token count (ties are broken in favor of one non-baseline condition). The Mean block at the bottom of the table reports the per-model average across all 14 tasks. The $\Delta$ block reports the change relative to the baseline (point difference for test accuracy, and percentage reduction for tokens).}
\label{tab:tuning-cls}
\begin{tabularx}{\textwidth}{cc l YY YY YY}
\toprule
\multirow{2}{*}{Database} & \multirow{2}{*}{Task} & \multirow{2}{*}{RFM} & \multicolumn{2}{c}{Baseline} & \multicolumn{2}{c}{STEER} & \multicolumn{2}{c}{STEER + Prob.\ Tuning} \\
\cmidrule(lr){4-5} \cmidrule(lr){6-7} \cmidrule(lr){8-9}
& & & Test $\uparrow$ & Tokens $\downarrow$ & Test $\uparrow$ & Tokens $\downarrow$ & Test $\uparrow$ & Tokens $\downarrow$ \\
\midrule
\multirow{2}{*}{rel-amazon} & \multirow{2}{*}{item-churn} & RT & 68.20 & 1023.2 & \textbf{71.06} & \cellcolor{green!15}230.2 & 71.06 & 230.2 \\
 &  & RT-J & 72.26 & 1018.9 & \textbf{74.21} & \cellcolor{green!15}118.7 & 74.21 & 118.7 \\
\midrule
\multirow{2}{*}{rel-amazon} & \multirow{2}{*}{user-churn} & RT & 64.14 & 1024.0 & \textbf{65.29} & 82.5 & 64.55 & \cellcolor{green!15}71.2 \\
 &  & RT-J & 65.45 & 1021.3 & \textbf{65.53} & 68.1 & 65.13 & \cellcolor{green!15}58.6 \\
\midrule
\multirow{2}{*}{rel-arxiv} & \multirow{2}{*}{paper-citation} & RT & 79.85 & 989.6 & \textbf{79.99} & \cellcolor{green!15}863.2 & 79.75 & 919.2 \\
 &  & RT-J & \textbf{80.54} & 997.3 & 79.99 & \cellcolor{green!15}862.7 & 80.33 & 922.3 \\
\midrule
\multirow{2}{*}{rel-avito} & \multirow{2}{*}{user-clicks} & RT & 51.92 & 613.6 & 63.80 & 487.5 & \textbf{66.07} & \cellcolor{green!15}149.5 \\
 &  & RT-J & \textbf{49.64} & 612.2 & 40.52 & 463.6 & 41.57 & \cellcolor{green!15}132.6 \\
\midrule
\multirow{2}{*}{rel-avito} & \multirow{2}{*}{user-visits} & RT & 55.98 & 706.7 & \textbf{59.47} & 374.8 & 59.47 & \cellcolor{green!15}336.1 \\
 &  & RT-J & 63.77 & 731.7 & \textbf{64.94} & 150.7 & 63.75 & \cellcolor{green!15}119.1 \\
\midrule
\multirow{2}{*}{rel-f1} & \multirow{2}{*}{driver-dnf} & RT & 78.55 & 1005.2 & \textbf{81.40} & 999.4 & 81.38 & \cellcolor{green!15}989.2 \\
 &  & RT-J & \textbf{78.50} & 1001.5 & 77.58 & 995.3 & 78.36 & \cellcolor{green!15}979.3 \\
\midrule
\multirow{2}{*}{rel-f1} & \multirow{2}{*}{driver-top3} & RT & 89.26 & 1000.2 & \textbf{90.70} & 1000.2 & 90.46 & \cellcolor{green!15}775.7 \\
 &  & RT-J & \textbf{89.23} & 996.6 & 89.15 & 996.8 & 88.96 & \cellcolor{green!15}337.6 \\
\midrule
\multirow{2}{*}{rel-hm} & \multirow{2}{*}{user-churn} & RT & 63.80 & 1024.0 & \textbf{65.40} & 283.3 & 65.26 & \cellcolor{green!15}130.2 \\
 &  & RT-J & \textbf{63.63} & 1020.3 & 61.49 & \cellcolor{green!15}197.6 & 62.45 & 994.2 \\
\midrule
\multirow{2}{*}{rel-ratebeer} & \multirow{2}{*}{beer-rating-churn} & RT & 68.25 & 1024.0 & 70.93 & 1005.8 & \textbf{71.38} & \cellcolor{green!15}897.2 \\
 &  & RT-J & 66.98 & 1011.7 & 74.82 & 976.7 & \textbf{75.81} & \cellcolor{green!15}875.5 \\
\midrule
\multirow{2}{*}{rel-ratebeer} & \multirow{2}{*}{brewer-dormant} & RT & \textbf{71.79} & 1024.0 & 69.54 & 621.4 & 70.59 & \cellcolor{green!15}613.0 \\
 &  & RT-J & 61.66 & 1007.8 & \textbf{61.83} & 631.8 & 61.41 & \cellcolor{green!15}496.5 \\
\midrule
\multirow{2}{*}{rel-ratebeer} & \multirow{2}{*}{user-rating-churn} & RT & 73.38 & 1024.0 & 73.02 & 887.3 & \textbf{82.75} & \cellcolor{green!15}361.4 \\
 &  & RT-J & 84.29 & 1013.1 & 85.72 & \cellcolor{green!15}898.2 & \textbf{87.37} & 1005.0 \\
\midrule
\multirow{2}{*}{rel-stack} & \multirow{2}{*}{user-badge} & RT & \textbf{81.69} & 355.6 & 75.89 & 134.2 & 73.67 & \cellcolor{green!15}39.5 \\
 &  & RT-J & \textbf{89.53} & 357.8 & 87.01 & \cellcolor{green!15}125.0 & 86.26 & 225.8 \\
\midrule
\multirow{2}{*}{rel-stack} & \multirow{2}{*}{user-engagement} & RT & 82.83 & 981.5 & 85.15 & 656.8 & \textbf{88.64} & \cellcolor{green!15}115.0 \\
 &  & RT-J & \textbf{87.59} & 975.0 & 87.52 & \cellcolor{green!15}737.3 & 85.54 & 847.1 \\
\midrule
\multirow{2}{*}{rel-trial} & \multirow{2}{*}{study-outcome} & RT & 50.05 & 1022.8 & 48.83 & \cellcolor{green!15}25.7 & \textbf{52.76} & 28.1 \\
 &  & RT-J & 51.08 & 1011.0 & \textbf{52.68} & \cellcolor{green!15}25.2 & 47.11 & 27.3 \\
\midrule
\multirow{2}{*}{Mean} & \multirow{2}{*}{(14 tasks)}
 & RT & 69.98 & 915.6 & 71.46 & 546.6 & \textbf{72.70} & \cellcolor{green!15}404.0 \\
 &  & RT-J & \textbf{71.72} & 912.6 & 71.64 & 517.7 & 71.30 & \cellcolor{green!15}510.0 \\
\midrule
\multirow{2}{*}{$\Delta$} & \multirow{2}{*}{vs.\ baseline}
 & RT & --- & --- & +1.48 & -40.3\% & +2.72 & -55.9\% \\
 &  & RT-J & --- & --- & -0.08 & -43.3\% & -0.42 & -44.1\% \\
\bottomrule
\end{tabularx}
\end{table*}

\begin{table*}[t]
\centering
\small
\caption{Probability tuning ablation: regression tasks ($R^2$~\%, higher is better). For each task and model, we report test accuracy and number of tokens with baseline sampling, STEER with the default probability profile, and STEER with a probability profile tuned for the specific task. Within each row, bold indicates the highest test score and the highlighted cell indicates the lowest token count (ties are broken in favor of one non-baseline condition). The Mean block at the bottom of the table reports the per-model average across all 10 tasks. The $\Delta$ block reports the change relative to the baseline (point difference for test accuracy, and percentage reduction for tokens).}
\label{tab:tuning-reg}
\begin{tabularx}{\textwidth}{cc l YY YY YY}
\toprule
\multirow{2}{*}{Database} & \multirow{2}{*}{Task} & \multirow{2}{*}{RFM} & \multicolumn{2}{c}{Baseline} & \multicolumn{2}{c}{STEER} & \multicolumn{2}{c}{STEER + Prob.\ Tuning} \\
\cmidrule(lr){4-5} \cmidrule(lr){6-7} \cmidrule(lr){8-9}
& & & Test $\uparrow$ & Tokens $\downarrow$ & Test $\uparrow$ & Tokens $\downarrow$ & Test $\uparrow$ & Tokens $\downarrow$ \\
\midrule
\multirow{2}{*}{rel-amazon} & \multirow{2}{*}{item-ltv} & RT & 4.59 & 957.2 & 4.26 & 203.5 & \textbf{36.05} & \cellcolor{green!15}174.3 \\
 &  & RT-J & 59.31 & 951.5 & \textbf{60.17} & \cellcolor{green!15}108.4 & 60.17 & 108.4 \\
\midrule
\multirow{2}{*}{rel-amazon} & \multirow{2}{*}{user-ltv} & RT & 15.20 & 1023.2 & \textbf{26.26} & 100.8 & 25.73 & \cellcolor{green!15}70.3 \\
 &  & RT-J & \textbf{61.52} & 1021.3 & 57.19 & 86.9 & 54.41 & \cellcolor{green!15}58.6 \\
\midrule
\multirow{2}{*}{rel-arxiv} & \multirow{2}{*}{author-publication} & RT & 22.77 & 740.5 & \textbf{24.75} & 491.4 & 22.57 & \cellcolor{green!15}46.9 \\
 &  & RT-J & 18.46 & 728.8 & \textbf{33.45} & \cellcolor{green!15}474.4 & 33.45 & 474.4 \\
\midrule
\multirow{2}{*}{rel-avito} & \multirow{2}{*}{ad-ctr} & RT & 4.19 & 1024.0 & 5.88 & 999.4 & \textbf{9.04} & \cellcolor{green!15}675.2 \\
 &  & RT-J & 2.44 & 1020.8 & -3.02 & 994.6 & \textbf{2.63} & \cellcolor{green!15}149.4 \\
\midrule
\multirow{2}{*}{rel-f1} & \multirow{2}{*}{driver-position} & RT & 45.31 & 982.6 & 44.69 & \cellcolor{green!15}982.6 & \textbf{46.01} & 982.6 \\
 &  & RT-J & 55.40 & 978.9 & 55.34 & 978.4 & \textbf{55.45} & \cellcolor{green!15}975.6 \\
\midrule
\multirow{2}{*}{rel-hm} & \multirow{2}{*}{item-sales} & RT & 21.19 & 751.5 & \textbf{23.74} & \cellcolor{green!15}313.4 & 23.74 & 313.4 \\
 &  & RT-J & 44.12 & 729.3 & \textbf{44.30} & \cellcolor{green!15}186.6 & 44.30 & 186.6 \\
\midrule
\multirow{2}{*}{rel-ratebeer} & \multirow{2}{*}{user-rating-count} & RT & 12.91 & 1024.0 & 19.47 & 420.2 & \textbf{21.74} & \cellcolor{green!15}350.9 \\
 &  & RT-J & 40.83 & 1013.0 & 45.06 & 272.6 & \textbf{46.15} & \cellcolor{green!15}178.3 \\
\midrule
\multirow{2}{*}{rel-stack} & \multirow{2}{*}{post-votes} & RT & 28.75 & 1003.0 & \textbf{29.18} & \cellcolor{green!15}413.5 & 29.18 & 942.6 \\
 &  & RT-J & \textbf{86.08} & 989.1 & 85.25 & \cellcolor{green!15}139.5 & 84.73 & 512.1 \\
\midrule
\multirow{2}{*}{rel-trial} & \multirow{2}{*}{site-success} & RT & 14.78 & 1015.3 & 34.83 & 944.9 & \textbf{34.96} & \cellcolor{green!15}935.0 \\
 &  & RT-J & 43.54 & 924.8 & \textbf{46.66} & 794.8 & 46.18 & \cellcolor{green!15}789.8 \\
\midrule
\multirow{2}{*}{rel-trial} & \multirow{2}{*}{study-adverse} & RT & \textbf{3.57} & 1024.0 & -0.92 & \cellcolor{green!15}21.7 & 1.38 & 410.8 \\
 &  & RT-J & -7.37 & 1014.1 & -12.19 & \cellcolor{green!15}22.0 & \textbf{-1.07} & 581.7 \\
\midrule
\multirow{2}{*}{Mean} & \multirow{2}{*}{(10 tasks)}
 & RT & 17.33 & 954.5 & 21.21 & \cellcolor{green!15}489.1 & \textbf{25.04} & 490.2 \\
 &  & RT-J & 40.43 & 937.2 & 41.22 & 405.8 & \textbf{42.64} & \cellcolor{green!15}401.5 \\
\midrule
\multirow{2}{*}{$\Delta$} & \multirow{2}{*}{vs.\ baseline}
 & RT & --- & --- & +3.89 & -48.8\% & +7.71 & -48.6\% \\
 &  & RT-J & --- & --- & +0.79 & -56.7\% & +2.21 & -57.2\% \\
\bottomrule
\end{tabularx}
\end{table*}

\vspace{-8pt}
\subsection{Ablation: Probability Tuning}
\label{sec:prob-tuning}

Throughout this paper, we have used a single, default
tier-to-probability mapping (which we also refer to as a probability profile).
This experiment 
tests whether tuning this mapping on a per-task basis, using a 
coordinate-ascent search over the validation split, improves upon this 
default. Because the tuned profile is chosen independently for every 
task, we report full per-task results in 
Tables~\ref{tab:tuning-cls} and~\ref{tab:tuning-reg} rather than 
averages alone.

Probability tuning finds a substantially different probability profile for nearly every task. Tables~\ref{tab:tuning-cls} and~\ref{tab:tuning-reg} show that the benefit of tuning varies widely. Tuning mostly reduces sample size, and it improves accuracy dramatically for some tasks: on 
rel-ratebeer/user-rating-churn, RT's AUROC rises from $73.02$ 
with the default profile to $82.75$ with tuning, and on rel-amazon/item-ltv, its $R^2$ rises  from $4.26$ to $36.05$.

On the other hand, the accuracy decreases with tuning for some tasks, and RT-J's classification tasks show 
this clearly: AUROC of rel-trial/study-outcome drops from $52.68$ with the default profile to $47.11$ once tuned, and rel-stack/user-badge 
drops from $87.01$ to $86.26$. This is 
consistent with a known risk of any validation-based search: it can 
find a profile that fits the validation split without generalizing well to the test split.

On average, probability tuning improves accuracy for three of the four model/task-type combinations. The only exception is RT-J on classification tasks, due to the two tasks mentioned above for which tuning reduces accuracy. Overall, probability tuning can yield meaningful improvements over the default probability profile, but these gains are task- and model-dependent and are not guaranteed. Moreover, tuning incurs additional cost because it requires searching over multiple probability profiles. Thus, while probability tuning is worthwhile when the potential gains justify this overhead, the main conclusion from this experiment is that the default probability profile provides a robust, effective, and inexpensive choice that performs well across models and tasks.

\vspace{-8pt}
\section{Related Work}
\label{sec:related-work}

\textbf{Relational Deep Learning.} Fey et al.~\cite{fey2024rdl} introduced an end-to-end framework for predictive modeling on relational databases, establishing relational deep learning (RDL) as a subfield of machine learning. At its core, RDL represents a database as a relational entity graph where each row is a node and every primary--foreign key relationship is an edge. Initial approaches applied heterogeneous graph neural networks directly to these graphs, and more recently advanced message-passing architectures have been proposed to improve expressiveness on relational data~\cite{chen2025relgnn}. Transformer-based architectures have also been explored as a more powerful alternative to GNN message passing, achieving stronger performance on relational benchmarks~\cite{peleska2025transformers, dwivedi2026relgt}. A fundamental limitation shared by all these approaches is that they are schema-specific: the model is tightly coupled to a particular database schema and cannot transfer to new, unseen databases.

\vspace{4pt}\noindent\textbf{Relational Foundation Models.} To overcome the schema-specificity of RDL, recent work has begun developing relational foundation models capable of zero-shot generalization across databases. KumoRFM and its successor KumoRFM-2~\cite{fey2025kumorfm1, hudovernik2026kumorfm2} propose a graph-transformer-based relational foundation model with both zero-shot and fine-tuning capabilities, but their implementation details and pretraining procedure are not publicly available. A separate line of work applies single-table tabular foundation models~\cite{hollmann2023tabpfn, hollmann2025tabpfn, qu2025tabicl, qu2026tabiclv2} to relational data by first flattening the database into a single feature table; RDBLearn~\cite{xu2026no} follows this strategy, aggregating relational neighborhoods into engineered features before applying an in-context tabular learner, an effective approach that nonetheless discards the relational structure a native model can exploit directly. Two architecturally distinct open-source relational foundation models represent the current state of the art. Griffin~\cite{wang2025griffin} pretrains a graph-centric architecture on a mixture of tabular and relational datasets, combining table-level encoders with graph neural network propagation across tables. The Relational Transformer (RT)~\cite{ranjan2026rt} instead represents every database cell as an individual token, introducing cell-level tokenization, task table prompting, and a novel Relational Attention mechanism to enable zero-shot inference across heterogeneous schemas. Two recent follow-ups retain this same cell-level architecture while changing how it is pretrained: RT-J~\cite{ranjan2026rtj} introduces a new large-scale pretraining recipe aimed at label-efficient few-shot prediction, and RT-PluRel~\cite{kothapalli2026plurel} pretrains on synthetic relational databases generated by the PluRel framework. One of the defining features of STEER is that it can work with any RFM since its decisions are at the level of which foreign-key link to follow during sampling. We demonstrated this flexibility by evaluating STEER on three RFMs with different architectures and different samplers.

\vspace{4pt}\noindent\textbf{Sampling Strategies for Graph-Based Models.} A rich body of work has developed neighborhood sampling methods for GNNs, primarily motivated by training efficiency. Hamilton et al.~\cite{hamilton2017graphsage} introduced uniform neighbor sampling as part of GraphSAGE, enabling scalable aggregation over sampled local neighborhoods. Ying et al.~\cite{ying2018pinsage} extended this idea with random-walk-based importance sampling in PinSage, biasing neighborhood selection toward more relevant nodes. Lao and Cohen~\cite{lao2010pathconstrained} similarly use path-constrained random walks to guide relational retrieval along specified relation paths. In the RDL setting, Khan and Aboulnaga~\cite{khan2025sqlrdl} explore SQL-based sampling within relational database systems as part of a broader vision for performing RDL computation inside the DBMS. More recently, Afandi et al.~\cite{afandi2026kgwise} proposed using LLMs to generate reusable query templates that extract semantically relevant subgraphs for GNN inference on large knowledge graphs, providing semantic guidance at inference time. Ding et al.~\cite{ding2026gelgt} introduce a structure-semantic collaborative sampling strategy for a task-specific relational graph transformer that filters irrelevant semantic information while preserving structural connectivity. Our work addresses a complementary setting by introducing a semantically informed traversal filter for inference-time sampling in a zero-shot relational foundation model, where no task-specific training data is available to learn what is informative.

\vspace{4pt}\noindent\textbf{LLMs for Relational Data.} Large language models have been applied to relational data in several ways. Wydmuch et al.~\cite{wydmuch2024llmrelational} demonstrate that even a straightforward application of LLMs to relational databases---by serializing the database subgraph into flat text---achieves competitive performance on RelBench tasks, establishing LLMs as a promising baseline. Wu et al.~\cite{wu2025relllm} improve on this by introducing Rel-LLM, a hybrid architecture that uses a GNN encoder to generate structured relational prompts for LLMs within a retrieval-augmented generation framework, preserving relational structure rather than serializing to flat text. Beyond end-to-end prediction, LLMs have also been shown to reason effectively about schema semantics from attribute names and descriptions alone, both in the context of schema matching~\cite{parciak2024schemamatching} and in text-to-SQL systems that link natural-language questions to the correct tables and columns purely from their names~\cite{li2024codes}. Our approach differs fundamentally from all of these prior approaches: rather than using LLMs to make predictions, match schemas, or decide how much schema to expose to another model, we use them as a semantic oracle that produces a tiered relevance ranking of a database schema's traversal edges to steer sampling at inference time.
\vspace{-16pt}
\section{Conclusion}
\label{sec:conclusion}

We presented STEER, a semantically informed sampling approach for reducing RFM inference cost. STEER uses an LLM to assess the relevance of schema edges to a prediction task and converts these judgments into traversal probabilities for sample construction. The resulting relevance configuration is reused across predictions, amortizing the cost of the LLM calls. Across three RFMs, STEER substantially reduces inference context size while preserving, and in several cases improving, prediction accuracy. In contrast, uninformed random pruning causes substantial accuracy degradation, showing that effective context reduction requires reasoning about which parts of the relational structure are relevant to the task.

Several directions remain for future work. First, integrating STEER into multiple RFMs highlights the value of a common SQL-based sampling abstraction that can be shared across models. Second, the optional column filter suggests that semantic guidance can extend beyond edge traversal, although its mixed results indicate that more robust methods are needed, potentially combining schema semantics with lightweight data statistics. Finally, task-specific probability profiles can sometimes outperform the default profile, motivating inexpensive methods for adapting traversal probabilities without costly search. More broadly, applying semantic guidance to other RFM architectures and sampling strategies may yield further reductions in inference cost without sacrificing predictive accuracy.

\bibliographystyle{ACM-Reference-Format}
\bibliography{references}

@inproceedings{fey2024rdl,
  author    = {Matthias Fey and Weihua Hu and Kexin Huang and Jan Eric Lenssen
               and Rishabh Ranjan and Joshua Robinson and Rex Ying and
               Jiaxuan You and Jure Leskovec},
  title     = {Position: Relational Deep Learning---{G}raph Representation
               Learning on Relational Databases},
  booktitle = {Proceedings of the International Conference on Machine
               Learning (ICML)},
  pages     = {13592--13607},
  year      = {2024},
}

@inproceedings{chen2025relgnn,
  author    = {Tianlang Chen and Charilaos Kanatsoulis and Jure Leskovec},
  title     = {{RelGNN}: Composite Message Passing for Relational Deep Learning},
  booktitle = {Proceedings of the International Conference on Machine
               Learning (ICML)},
  pages     = {8296--8312},
  year      = {2025},
}

@article{peleska2025transformers,
  author  = {Jakub Peleska and Gustav {S\'{\i}r}},
  title   = {Tabular Transformers Meet Relational Databases},
  journal = {{ACM} Transactions on Intelligent Systems and Technology (TIST)},
  volume  = {16},
  number  = {5},
  pages   = {115:1--115:24},
  year    = {2025},
}

@inproceedings{dwivedi2026relgt,
  author    = {Vijay Prakash Dwivedi and Sri Jaladi and Yangyi Shen and
               Federico Lopez and Charilaos Kanatsoulis and Rishi Puri and
               Matthias Fey and Jure Leskovec},
  title     = {Relational Graph Transformer},
  booktitle = {Proceedings of the International Conference on Learning
               Representations (ICLR)},
  pages     = {154348--154369},
  year      = {2026},
}

@inproceedings{robinson2024relbench,
  author    = {Joshua Robinson and Rishabh Ranjan and Weihua Hu and
               Kexin Huang and Jiaqi Han and Alejandro Dobles and
               Matthias Fey and Jan Eric Lenssen and Yiwen Yuan and
               Zecheng Zhang and Jure Leskovec},
  title     = {{RelBench}: {A} Benchmark for Deep Learning on Relational
               Databases},
  booktitle = {Advances in Neural Information Processing Systems 37 (NeurIPS)},
  year      = {2024},
}

@inproceedings{gu2026relbenchv2,
  author    = {Justin Gu and Rishabh Ranjan and Charilaos I. Kanatsoulis and
               Haiming Tang and Martin Jurkovic and Valter Hudovernik and
               Mark Znidar and Pranshu Chaturvedi and Parth Shroff and
               Fengyu Li and Jure Leskovec},
  title     = {{RelBench} v2: {A} Large-Scale Benchmark and Repository for
               Relational Data},
  booktitle = {Proceedings of the International Conference on Learning
               Representations (ICLR) Workshop on Foundation Models for
               Tabular and Structured Data},
  year      = {2026},
}

@inproceedings{wang2025griffin,
  author    = {Yanbo Wang and Xiyuan Wang and Quan Gan and Minjie Wang and
               Qibin Yang and David Wipf and Muhan Zhang},
  title     = {Griffin: Towards a Graph-Centric Relational Database Foundation
               Model},
  booktitle = {Proceedings of the 42nd International Conference on Machine
               Learning (ICML)},
  year      = {2025},
}

@techreport{fey2025kumorfm1,
  author      = {Matthias Fey and Vid Kocijan and Federico L{\'o}pez and
                 Jan Eric Lenssen and Jure Leskovec},
  title       = {{KumoRFM}: A Foundation Model for In-Context Learning
                 on Relational Data},
  institution = {Kumo AI},
  year        = {2025},
}

@article{hudovernik2026kumorfm2,
  author  = {Valter Hudovernik and Federico Lopez and Vid Kocijan and
             Akihiro Nitta and Jan Eric Lenssen and Jure Leskovec and
             Matthias Fey},
  title   = {{KumoRFM}-2: Scaling Foundation Models for Relational
             Learning},
  journal = {CoRR},
  volume  = {abs/2604.12596},
  year    = {2026},
}

@inproceedings{ranjan2026rt,
  author    = {Rishabh Ranjan and Valter Hudovernik and Mark Znidar and
               Charilaos I. Kanatsoulis and Roshan Upendra and
               Mahmoud Mohammadi and Joe Meyer and Tom Palczewski and
               Carlos Guestrin and Jure Leskovec},
  title     = {Relational Transformer: Toward Zero-Shot Foundation Models for
               Relational Data},
  booktitle = {Proceedings of the International Conference on Learning
               Representations (ICLR)},
  year      = {2026},
}

@inproceedings{kothapalli2026plurel,
  author    = {Vignesh Kothapalli and Rishabh Ranjan and Valter Hudovernik and
               Vijay Prakash Dwivedi and Johannes Hoffart and Carlos Guestrin and
               Jure Leskovec},
  title     = {{PluRel}: Synthetic Data Unlocks Scaling Laws for Relational
               Foundation Models},
  booktitle = {Proceedings of the Forty-Third International Conference on
               Machine Learning (ICML)},
  year      = {2026},
}

@inproceedings{ranjan2026rtj,
  author    = {Rishabh Ranjan and Vignesh Kothapalli and Harshvardhan Agarwal
               and Charilaos I. Kanatsoulis and Roshan Reddy Upendra and
               Tom Palczewski and Carlos Guestrin and Jure Leskovec},
  title     = {Large-Scale Pretraining Unlocks Few-Shot Prediction for
               Relational Data},
  booktitle = {Proceedings of the 2nd Workshop on Foundation Models for
               Structured Data at the International Conference on Machine
               Learning (ICML)},
  year      = {2026},
}

@inproceedings{xu2026no,
  author    = {Linjie Xu and Yanlin Zhang and Quan Gan and
               Minjie Wang and David Wipf},
  title     = {No Need to Train Your {RDB} Foundation Model},
  booktitle = {Proceedings of the Forty-Third International Conference on
               Machine Learning (ICML)},
  year      = {2026},
}

@inproceedings{kocijan2026pql,
  author    = {Vid Kocijan and Jinu Sunil and Jan Eric Lenssen and
               Viman Deb and Xinwei He and Federico Reyes Gomez and
               Matthias Fey and Jure Leskovec},
  title     = {{Predictive Query Language}: {A} Domain-Specific Language for
               Predictive Modeling on Relational Databases},
  booktitle = {Proceedings of the International Workshop on Tabular Data
               Analysis (TaDA), co-located with the International Conference
               on Very Large Data Bases (VLDB)},
  year      = {2026},
}

@inproceedings{hamilton2017graphsage,
  author    = {William L. Hamilton and Zhitao Ying and Jure Leskovec},
  title     = {Inductive Representation Learning on Large Graphs},
  booktitle = {Advances in Neural Information Processing Systems 30 (NeurIPS)},
  pages     = {1024--1034},
  year      = {2017},
}

@inproceedings{ying2018pinsage,
  author    = {Rex Ying and Ruining He and Kaifeng Chen and Pong Eksombatchai
               and William L. Hamilton and Jure Leskovec},
  title     = {Graph Convolutional Neural Networks for Web-Scale Recommender
               Systems},
  booktitle = {Proceedings of the 24th {ACM} {SIGKDD} International Conference
               on Knowledge Discovery and Data Mining (KDD)},
  pages     = {974--983},
  year      = {2018},
}

@inproceedings{afandi2026kgwise,
  author    = {Waleed Afandi and Hussein Abdallah and
               Ashraf Aboulnaga and Essam Mansour},
  title     = {An {LLM}-Guided Query-Aware Inference System for {GNN} Models
               on Large Knowledge Graphs},
  booktitle = {Proceedings of the 42nd {IEEE} International Conference on
               Data Engineering (ICDE)},
  pages     = {1675--1688},
  year      = {2026},
}

@article{ding2026gelgt,
  author  = {Zezhong Ding and Jin Li and Xugang Wang and Xike Xie},
  title   = {Gaussian Relational Graph Transformer},
  journal = {CoRR},
  volume  = {abs/2605.15575},
  year    = {2026},
}

@article{lao2010pathconstrained,
  author  = {Ni Lao and William W. Cohen},
  title   = {Relational Retrieval Using a Combination of
             Path-Constrained Random Walks},
  journal = {Machine Learning},
  year    = {2010},
}

@inproceedings{khan2025sqlrdl,
  author    = {Fahim Shahriar Khan and Ashraf Aboulnaga},
  title     = {A Vision for {SQL}-Based Relational Deep Learning},
  booktitle = {Proceedings of the International Workshop on Tabular Data
               Analysis (TaDA), co-located with the International Conference
               on Very Large Data Bases (VLDB)},
  year      = {2025},
}

@article{wydmuch2024llmrelational,
  author  = {Marek Wydmuch and {\L}ukasz Borchmann and Filip Grali{\'n}ski},
  title   = {Tackling Prediction Tasks in Relational Databases with {LLM}s},
  journal = {CoRR},
  volume  = {abs/2411.11829},
  year    = {2024},
}

@inproceedings{wu2025relllm,
  author    = {Fang Wu and Vijay Prakash Dwivedi and Jure Leskovec},
  title     = {Large Language Models are Good Relational Learners},
  booktitle = {Proceedings of the 63rd Annual Meeting of the Association for
               Computational Linguistics (ACL), Volume 1: Long Papers},
  pages     = {7835--7854},
  year      = {2025},
}

@inproceedings{parciak2024schemamatching,
  author    = {Marcel Parciak and Brecht Vandevoort and Frank Neven and
               Liesbet M. Peeters and Stijn Vansummeren},
  title     = {Schema Matching with Large Language Models: an Experimental
               Study},
  booktitle = {Proceedings of Workshops at the 50th International Conference
               on Very Large Data Bases (VLDB)},
  year      = {2024},
}

@article{li2024codes,
  author  = {Haoyang Li and Jing Zhang and Hanbing Liu and Ju Fan and
             Xiaokang Zhang and Jun Zhu and Renjie Wei and Hongyan Pan
             and Cuiping Li and Hong Chen},
  title   = {{CodeS}: Towards Building Open-Source Language Models
             for Text-to-{SQL}},
  journal = {Proceedings of the {ACM} on Management of Data},
  volume  = {2},
  number  = {3},
  pages   = {1--28},
  year    = {2024},
}

@inproceedings{wei2022cot,
  author    = {Jason Wei and Xuezhi Wang and Dale Schuurmans and
               Maarten Bosma and Brian Ichter and Fei Xia and
               Ed H. Chi and Quoc V. Le and Denny Zhou},
  title     = {Chain-of-Thought Prompting Elicits Reasoning in Large Language
               Models},
  booktitle = {Advances in Neural Information Processing Systems 35 (NeurIPS)},
  year      = {2022},
}

@inproceedings{qin2024pairwise,
  author    = {Zhen Qin and Rolf Jagerman and Kai Hui and Honglei
               Zhuang and Junru Wu and Le Yan and Jiaming Shen and
               Tianqi Liu and Jialu Liu and Donald Metzler and
               Xuanhui Wang and Michael Bendersky},
  title     = {Large Language Models are Effective Text Rankers with
               Pairwise Ranking Prompting},
  booktitle = {Findings of the Association for Computational Linguistics:
               North American Chapter of the Association for Computational
               Linguistics (NAACL) 2024},
  pages     = {1504--1518},
  year      = {2024},
}

@article{marchesin2026stratification,
  author  = {Stefano Marchesin and Matteo Ceccarello and Gianmaria
             Silvello},
  title   = {{LLMs} as Stratification Signals for {KG} Accuracy
             Evaluation},
  journal = {Proceedings of the VLDB Endowment},
  volume  = {19},
  number  = {9},
  pages   = {2005--2018},
  year    = {2026},
}

@inproceedings{chen2016xgboost,
  author    = {Tianqi Chen and Carlos Guestrin},
  title     = {{XGBoost}: A Scalable Tree Boosting System},
  booktitle = {Proceedings of the 22nd {ACM} {SIGKDD} International
               Conference on Knowledge Discovery and Data Mining (KDD)},
  pages     = {785--794},
  year      = {2016},
}

@article{shwartzziv2022tabular,
  author  = {Ravid Shwartz-Ziv and Amitai Armon},
  title   = {Tabular Data: Deep Learning is not all you need},
  journal = {Information Fusion},
  volume  = {81},
  pages   = {84--90},
  year    = {2022},
}

@inproceedings{hollmann2023tabpfn,
  author    = {Noah Hollmann and Samuel M\"uller and Katharina
               Eggensperger and Frank Hutter},
  title     = {{TabPFN}: A Transformer That Solves Small Tabular
               Classification Problems in a Second},
  booktitle = {Proceedings of the Eleventh International Conference on
               Learning Representations (ICLR)},
  year      = {2023},
}

@article{hollmann2025tabpfn,
  author  = {Noah Hollmann and Samuel M\"uller and Lennart Purucker
             and Arjun Krishnakumar and Max K\"orfer and Shi Bin Hoo
             and Robin Tibor Schirrmeister and Frank Hutter},
  title   = {Accurate Predictions on Small Data with a Tabular
             Foundation Model},
  journal = {Nature},
  volume  = {637},
  number  = {8045},
  pages   = {319--326},
  year    = {2025},
}

@inproceedings{qu2025tabicl,
  author    = {Jingang Qu and David Holzm\"uller and Ga\"el Varoquaux and
               Marine Le Morvan},
  title     = {{TabICL}: A Tabular Foundation Model for In-Context
               Learning on Large Data},
  booktitle = {Proceedings of the 42nd International Conference on Machine
               Learning (ICML)},
  pages     = {50817--50847},
  year      = {2025},
}

@inproceedings{qu2026tabiclv2,
  author    = {Jingang Qu and David Holzm{\"u}ller and
               Ga{\"e}l Varoquaux and Marine Le Morvan},
  title     = {{TabICLv2}: {A} Better, Faster, Scalable, and Open
               Tabular Foundation Model},
  booktitle = {Proceedings of the Forty-Third International Conference on
               Machine Learning (ICML)},
  year      = {2026},
}

@article{zhou2024survey,
  title   = {A Survey on Efficient Inference for Large Language Models},
  author  = {Zhou, Zixuan and Ning, Xuefei and Hong, Ke and Fu, Tianyu
             and Xu, Jiaming and Li, Shiyao and Lou, Yuming and Wang, Luning
             and Yuan, Zhihang and Li, Xiuhong and Yan, Shengen and Dai, Guohao
             and Zhang, Xiao-Ping and Dong, Yuhan and Wang, Yu},
  journal = {CoRR},
  volume  = {abs/2404.14294},
  year    = {2024},
}

@inproceedings{li2023compressing,
  title     = {Compressing Context to Enhance Inference Efficiency of Large
               Language Models},
  author    = {Li, Yucheng and Dong, Bo and Guerin, Frank and Lin, Chenghua},
  booktitle = {Proceedings of the Conference on Empirical Methods in Natural
               Language Processing (EMNLP)},
  pages     = {6342--6353},
  year      = {2023},
}

@inproceedings{hooper2025squeezed,
  title     = {Squeezed Attention: Accelerating Long Context Length {LLM}
               Inference},
  author    = {Hooper, Coleman Richard Charles and Kim, Sehoon
               and Mohammadzadeh, Hiva and Maheswaran, Monishwaran
               and Zhao, Sebastian and Paik, June and Mahoney, Michael W.
               and Keutzer, Kurt and Gholami, Amir},
  booktitle = {Proceedings of the 63rd Annual Meeting of the Association for
               Computational Linguistics (ACL)},
  pages     = {32631--32652},
  year      = {2025},
}

@inproceedings{abts2020thinkfast,
  title     = {Think Fast: A Tensor Streaming Processor ({TSP}) for
               Accelerating Deep Learning Workloads},
  author    = {Abts, Dennis and Ross, Jonathan and Sparling, Jonathan
               and Wong-VanHaren, Mark and Baker, Max and Hawkins, Tom
               and Bell, Andrew and Thompson, John and Kahsai, Temesghen
               and Kimmell, Garrin and Hwang, Jennifer and Leslie-Hurd, Rebekah
               and Bye, Michael and Creswick, E. R. and Boyd, Matthew
               and Venigalla, Mahitha and Laforge, Evan and Purdy, Jon
               and Kamath, Purushotham and Maheshwari, Dinesh
               and Beidler, Michael and Rosseel, Geert and Ahmad, Omar
               and Gagarin, Gleb and Czekalski, Richard and Rane, Ashay
               and Parmar, Sahil and Werner, Jeff and Sproch, Jim
               and Macias, Adrian and Kurtz, Brian},
  booktitle = {Proceedings of the {ACM}/{IEEE} Annual International Symposium
               on Computer Architecture (ISCA)},
  pages     = {145--158},
  year      = {2020},
}

@misc{awsinferentia,
  author       = {{Amazon Web Services}},
  title        = {{AWS Inferentia}: High-Performance Machine Learning
                  Inference Accelerators},
  howpublished = {\url{https://aws.amazon.com/ai/machine-learning/inferentia/}},
}

@inproceedings{xiao2023smoothquant,
  title     = {SmoothQuant: Accurate and Efficient Post-Training Quantization
               for Large Language Models},
  author    = {Xiao, Guangxuan and Lin, Ji and Seznec, Mickael and Wu, Hao
               and Demouth, Julien and Han, Song},
  booktitle = {Proceedings of the 40th International Conference on Machine
               Learning (ICML)},
  year      = {2023},
}

@inproceedings{leviathan2023speculative,
  title     = {Fast Inference from Transformers via Speculative Decoding},
  author    = {Leviathan, Yaniv and Kalman, Matan and Matias, Yossi},
  booktitle = {Proceedings of the 40th International Conference on Machine
               Learning (ICML)},
  year      = {2023},
}

\clearpage
\appendix

\section{Prompt Templates}
\label{sec:appendix-prompts}

For reproducibility, we report the full text of the two prompt 
templates used to produce tiered rankings: the smart prompting 
strategy and the simple prompting baseline 
(\Cref{sec:prompting-strategy}).
When creatnig a configuration, each template is used twice for the two navigation directions, 
once for \PtoF{} and once for \FtoP{}.
Each 
template is shown in its \PtoF{} form in full; the \FtoP{} form is 
identical except where noted. Every prompt ends with a task-specific 
block, shown here as a placeholder, containing the task name, dataset, 
task table, task table columns, and the full database definition.

\vspace{-48pt}
\subsection{Smart Prompting Template}

\begin{Verbatim}[breaklines=true, fontsize=\small]
You are an expert in relational databases and machine learning on tabular data.

Your job is to rank traversal edges into tiers for a probabilistic BFS-like primary-to-foreign-key traversal policy based on semantic reasoning about the database schema.
In a relational database, a primary-to-foreign-key relationship means that a table with a primary key is referenced by other tables via foreign keys. Those referencing tables are the child tables. When the sampler is at a row in any table, it can expand outward into child tables by following those foreign key references. Each such expansion is a traversal edge -- a directed connection from the current parent table to a specific child table via a foreign key. Your goal is to rank each traversal edge into tiers based on how useful it is to follow that specific edge during sampling. The coin toss happens at the edge level when a node is being pushed into the frontier, so rejected edges are never enqueued at all.
Your ranking should be based purely on semantic reasoning: what each table represents, what its columns describe, how directly that information relates to the prediction target and the entity being predicted for, and crucially, what the source table context is when the edge is crossed.

Follow these steps in order:

Step 1 - Understand the database structure: Read through all the tables and their columns. For each table, form a clear understanding of what real-world entity or concept that table represents and what kind of information each column holds. Identify all primary keys and foreign keys and understand how the tables are connected to each other.

Step 2 - Understand the task: Identify the entity being predicted for, the target value being predicted, and what the prediction task is asking you to estimate or classify.

Step 3 - Build the edge list: Go through each table in the database and identify which tables have at least one foreign key defined. For each foreign key column in each such table, identify its parent table -- the table whose primary key is referenced by that foreign key column. Write out all traversal edges as source->destination:fk_col triples, where source is the parent table (where the sampler currently is), destination is the child table (where the sampler would traverse to), and fk_col is the exact foreign key column name defined on the destination table. Note that the arrow in source->destination represents the direction the sampler travels, not the direction of the foreign key. The foreign key is always defined on the destination table pointing back to the source. Each foreign key column in the database produces exactly one edge. If a destination table has multiple foreign keys pointing to the same source, each FK column produces a separate edge with a distinct fk_col suffix. Tables that have no foreign keys produce no edges as destinations. List every edge explicitly before moving on.

Step 4 - Rank edges by semantic relevance: Go through each edge in your edge list and reason about how useful it is to follow that specific edge during sampling. The key question for each edge is: given that the sampler is currently at a row in the source table, does crossing this edge to the destination table provide meaningful signal toward the prediction target and the entity being predicted for? Also consider what subset of the destination table's rows you would actually reach by following this edge from this specific source. A more targeted subset that directly characterizes the prediction entity is more valuable than a broad or noisy one. Two edges that share the same destination table can have very different values if their source tables provide different context -- reason about each edge independently.

Rank all edges into the following tiers: very_strong, strong, moderate, weak.

very_strong: following this edge from this specific source will almost always lead to rows that directly describe or characterize the prediction entity or its outcomes in a way that is immediately relevant to the target.
strong: following this edge provides important contextual signal closely related to the target, even if not directly causal, given the source context.
moderate: following this edge has some plausible relationship to the target but the connection requires more than one step of reasoning or is indirect given the source context.
weak: following this edge leads to rows that describe entities or events with no plausible direct or indirect connection to the prediction target given where the sampler currently is.

Step 5 - Ensure there is at least one edge in your output. If none qualify, go back to step 4 and select the single most relevant edge into very_strong.

Hard constraints: only use edges from your step-3 list (do not invent edges); use exact table/column names from the database definition; each edge appears in exactly one tier.

Return only a JSON object: {"config": {...tiers...}, "reasoning": {"step_1": "...", ..., "step_5": "..."}}.

[Task name / Dataset / Task table / Task table columns / Database definition inserted here]
\end{Verbatim}

The \FtoP{} variant of this template is identical except for 
the following. The opening direction is reframed: \textit{``a 
foreign-to-primary-key relationship means that a table holds a foreign 
key that references a primary key in another table\ldots the sampler 
can expand outward into parent tables.''} In Step 3, \textit{source} 
is the child table (the sampler's current position), \textit{destination} 
is the parent table, and \textit{fk\_col} is defined on the source table 
rather than the destination; self-referential edges, where a table's 
foreign key references its own primary key, are explicitly valid and 
must be included. In Step 4, the guidance on targeted subsets is 
reframed around the destination as a single parent entity: \textit{``a 
parent row that directly identifies or characterizes the prediction 
entity is more valuable than one that only provides generic or loosely 
related context.''} The tier definitions, Step 5, hard constraints, 
output format, and trailing task-context block are otherwise identical 
to the \PtoF{} version.

\subsection{Simple Prompting Template}

\begin{Verbatim}[breaklines=true, fontsize=\small]
You are an expert in relational databases and machine learning on tabular data.

Your job is to identify which traversal edges are useful for a probabilistic BFS-like primary-to-foreign-key traversal policy, based on the database schema and the prediction task below.
In a relational database, a primary-to-foreign-key relationship means that a table with a primary key is referenced by other tables via foreign keys. Those referencing tables are the child tables. When the sampler is at a row in any table, it can expand outward into child tables by following those foreign key references. Each such expansion is a traversal edge -- a directed connection from the current parent table to a specific child table via a foreign key, written as source->destination:fk_col, where source is the parent table (where the sampler currently is) and destination is the child table (where the sampler would traverse to).

List the traversal edges you consider useful for predicting the target -- edges that lead to rows providing meaningful signal about the prediction target and the entity being predicted for.

Use the exact table names and column names from the database definition. Do not invent edges.

Return only a valid JSON object in this exact format, with no explanation, no markdown, and nothing else:
{"edges": ["source_a->destination_a:fk_col_x", "source_b->destination_b:fk_col_y"]}

[Task name / Dataset / Task table / Task table columns / Database definition inserted here]
\end{Verbatim}

The \FtoP{} variant reframes the opening direction the same 
way as the smart prompt's \FtoP{} variant: \textit{``a 
foreign-to-primary-key relationship means that a table holds a foreign 
key that references a primary key in another table\ldots the sampler 
can expand outward into parent tables\ldots source is the child table, 
destination is the parent table.''} Self-referential edges are again 
explicitly valid. The instruction line, hard constraints, JSON output 
format, and trailing task-context block are otherwise identical to the 
\PtoF{} version.

Every edge returned by a simple prompt is placed entirely into the 
\textit{very\_strong} tier, with the remaining three tiers left empty; 
the resulting configuration is then evaluated using a profile that 
zeroes every tier except \textit{very\_strong}. This is the mechanical 
difference between smart and simple prompting: a graduated, four-tier 
ranking produced through explicit step-by-step reasoning, versus a 
binary include-or-exclude decision with no reasoning scaffold.

\makeatletter
\let\@vspace\@vspace@orig
\let\@vspacer\@vspacer@orig
\makeatother
\section{Full Per-Task Results}
\label{sec:appendix-full-results}

We report the full per-task results for every experiment in
Section~\ref{sec:experiments}. All values are averaged using the same
evaluation procedure described in Section~\ref{sec:experimental-setup}.

\begin{table*}[p]
\centering
\footnotesize
\caption{Main results: classification tasks (AUROC~\%, higher is better). Used in \Cref{fig:main-cls}. For each task and model, we report test accuracy and number of tokens with the baseline (random sampling), naive pruning (full-budget sampling, then half the resulting sample discarded at random), STEER with Claude-based ranking, and STEER with GPT-based ranking. Within each row, bold indicates the highest test score and the highlighted cell indicates the lowest token count (ties are broken in favor of one non-baseline condition). The Mean block at the bottom of the table reports the per-model average across all 14 tasks. The $\Delta$ block reports the change relative to the baseline (point difference for test accuracy, and percentage reduction for tokens).}
\label{tab:appendix-main-cls}
\begin{tabularx}{\textwidth}{ccl YY YY YY YY}
\toprule
\multirow{2}{*}{Database} & \multirow{2}{*}{Task} & \multirow{2}{*}{RFM} & \multicolumn{2}{c}{Baseline} & \multicolumn{2}{c}{Naive Pruning} & \multicolumn{2}{c}{STEER (Claude)} & \multicolumn{2}{c}{STEER (GPT)} \\
\cmidrule(lr){4-5} \cmidrule(lr){6-7} \cmidrule(lr){8-9} \cmidrule(lr){10-11}
& & & Test $\uparrow$ & Tokens $\downarrow$ & Test $\uparrow$ & Tokens $\downarrow$ & Test $\uparrow$ & Tokens $\downarrow$ & Test $\uparrow$ & Tokens $\downarrow$ \\
\midrule
\multirow{3}{*}{rel-amazon} & \multirow{3}{*}{item-churn} & RT & 68.20 & 1023.2 & 63.90 & 511.6 & 71.06 & \cellcolor{green!15}230.2 & \textbf{71.07} & 236.4 \\
 &  & RT-J & 72.26 & 1018.9 & 66.58 & 509.9 & 74.21 & \cellcolor{green!15}118.7 & \textbf{74.41} & 122.6 \\
 &  & Griffin & 69.04 & 10885.8 & 65.64 & \cellcolor{green!15}5523.0 & \textbf{69.10} & 8601.3 & 69.07 & 9057.0 \\
\midrule
\multirow{3}{*}{rel-amazon} & \multirow{3}{*}{user-churn} & RT & 64.14 & 1024.0 & 60.39 & 512.0 & \textbf{65.29} & 82.5 & 64.55 & \cellcolor{green!15}71.2 \\
 &  & RT-J & 65.45 & 1021.3 & 62.09 & 511.2 & \textbf{65.53} & 68.1 & 65.05 & \cellcolor{green!15}58.6 \\
 &  & Griffin & 62.18 & 6111.9 & 60.73 & \cellcolor{green!15}3135.9 & 61.93 & 5016.0 & \textbf{62.71} & 3370.5 \\
\midrule
\multirow{3}{*}{rel-arxiv} & \multirow{3}{*}{paper-citation} & RT & 79.85 & 989.6 & 73.98 & \cellcolor{green!15}494.8 & \textbf{79.99} & 863.2 & 79.39 & 945.0 \\
 &  & RT-J & 80.54 & 997.3 & 74.26 & \cellcolor{green!15}499.2 & 79.99 & 862.7 & \textbf{80.81} & 947.5 \\
 &  & Griffin & \textbf{74.49} & 5488.9 & 71.13 & \cellcolor{green!15}2800.4 & 74.04 & 3234.5 & 74.47 & 3363.1 \\
\midrule
\multirow{3}{*}{rel-avito} & \multirow{3}{*}{user-clicks} & RT & 51.92 & 613.6 & 53.45 & \cellcolor{green!15}307.2 & \textbf{63.80} & 487.5 & 63.02 & 534.4 \\
 &  & RT-J & \textbf{49.64} & 612.2 & 44.99 & \cellcolor{green!15}306.6 & 40.52 & 463.6 & 45.30 & 521.9 \\
 &  & Griffin & 50.44 & 14308.2 & 49.31 & \cellcolor{green!15}7238.0 & 50.41 & 8676.1 & \textbf{50.47} & 9835.4 \\
\midrule
\multirow{3}{*}{rel-avito} & \multirow{3}{*}{user-visits} & RT & 55.98 & 706.7 & 57.14 & \cellcolor{green!15}353.6 & 59.47 & 374.8 & \textbf{59.57} & 603.8 \\
 &  & RT-J & 63.77 & 731.7 & 57.02 & 366.4 & 64.94 & \cellcolor{green!15}150.7 & \textbf{65.07} & 580.4 \\
 &  & Griffin & 59.94 & 17505.3 & 59.86 & \cellcolor{green!15}8836.3 & 60.06 & 12793.5 & \textbf{60.16} & 9244.0 \\
\midrule
\multirow{3}{*}{rel-f1} & \multirow{3}{*}{driver-dnf} & RT & 78.55 & 1005.2 & 73.16 & \cellcolor{green!15}502.6 & \textbf{81.40} & 999.4 & 80.62 & 1000.8 \\
 &  & RT-J & \textbf{78.50} & 1001.5 & 74.27 & \cellcolor{green!15}501.3 & 77.58 & 995.3 & 78.44 & 996.9 \\
 &  & Griffin & 51.20 & 44371.9 & 51.65 & \cellcolor{green!15}22308.0 & 48.69 & 26472.8 & \textbf{51.84} & 23747.1 \\
\midrule
\multirow{3}{*}{rel-f1} & \multirow{3}{*}{driver-top3} & RT & 89.26 & 1000.2 & 82.90 & \cellcolor{green!15}500.1 & \textbf{90.70} & 1000.2 & 89.35 & 1000.2 \\
 &  & RT-J & \textbf{89.23} & 996.6 & 85.50 & \cellcolor{green!15}498.8 & 89.15 & 996.8 & 89.18 & 996.4 \\
 &  & Griffin & 61.75 & 44002.5 & 60.42 & \cellcolor{green!15}22117.1 & 60.10 & 32050.6 & \textbf{61.88} & 37211.2 \\
\midrule
\multirow{3}{*}{rel-hm} & \multirow{3}{*}{user-churn} & RT & 63.80 & 1024.0 & 61.61 & 512.0 & \textbf{65.40} & \cellcolor{green!15}283.3 & 65.40 & 283.3 \\
 &  & RT-J & \textbf{63.63} & 1020.3 & 59.38 & 510.7 & 61.49 & \cellcolor{green!15}197.6 & 61.49 & 197.6 \\
 &  & Griffin & \textbf{60.38} & 23628.7 & 59.97 & \cellcolor{green!15}12012.1 & 60.05 & 18939.4 & 60.05 & 18939.4 \\
\midrule
\multirow{3}{*}{rel-ratebeer} & \multirow{3}{*}{beer-rating-churn} & RT & 68.25 & 1024.0 & 65.93 & \cellcolor{green!15}512.0 & \textbf{70.93} & 1005.8 & 70.25 & 1024.0 \\
 &  & RT-J & 66.98 & 1011.7 & 64.66 & \cellcolor{green!15}506.4 & 74.82 & 976.7 & \textbf{75.75} & 1009.6 \\
 &  & Griffin & \textbf{62.78} & 46983.0 & 61.34 & 23765.8 & 62.28 & \cellcolor{green!15}15021.2 & 61.84 & 18660.7 \\
\midrule
\multirow{3}{*}{rel-ratebeer} & \multirow{3}{*}{brewer-dormant} & RT & \textbf{71.79} & 1024.0 & 67.57 & \cellcolor{green!15}512.0 & 69.54 & 621.4 & 70.01 & 1024.0 \\
 &  & RT-J & 61.66 & 1007.8 & 55.40 & \cellcolor{green!15}504.4 & 61.83 & 631.8 & \textbf{63.96} & 1010.3 \\
 &  & Griffin & \textbf{68.74} & 145954.7 & 64.77 & 73251.0 & 67.91 & \cellcolor{green!15}41201.3 & 68.59 & 46707.2 \\
\midrule
\multirow{3}{*}{rel-ratebeer} & \multirow{3}{*}{user-rating-churn} & RT & \textbf{73.38} & 1024.0 & 72.08 & \cellcolor{green!15}512.0 & 73.02 & 887.3 & 72.37 & 887.4 \\
 &  & RT-J & 84.29 & 1013.1 & 83.17 & \cellcolor{green!15}507.1 & 85.72 & 898.2 & \textbf{86.10} & 896.6 \\
 &  & Griffin & 84.61 & 32913.9 & 84.52 & \cellcolor{green!15}16731.8 & 84.61 & 28994.1 & \textbf{84.63} & 28485.4 \\
\midrule
\multirow{3}{*}{rel-stack} & \multirow{3}{*}{user-badge} & RT & \textbf{81.69} & 355.6 & 75.95 & 178.1 & 75.89 & \cellcolor{green!15}134.2 & 78.77 & 172.7 \\
 &  & RT-J & \textbf{89.53} & 357.8 & 85.35 & 179.4 & 87.01 & \cellcolor{green!15}125.0 & 87.50 & 177.2 \\
 &  & Griffin & 80.69 & 10608.3 & 78.86 & \cellcolor{green!15}5418.5 & \textbf{80.73} & 7830.8 & 80.37 & 7558.6 \\
\midrule
\multirow{3}{*}{rel-stack} & \multirow{3}{*}{user-engagement} & RT & 82.83 & 981.5 & 71.42 & 490.8 & 85.15 & 656.8 & \textbf{85.80} & \cellcolor{green!15}270.7 \\
 &  & RT-J & \textbf{87.59} & 975.0 & 79.24 & 488.0 & 87.52 & 737.3 & 86.81 & \cellcolor{green!15}209.9 \\
 &  & Griffin & \textbf{76.78} & 23592.3 & 65.38 & \cellcolor{green!15}11910.7 & 75.69 & 18442.1 & 74.56 & 12756.7 \\
\midrule
\multirow{3}{*}{rel-trial} & \multirow{3}{*}{study-outcome} & RT & 50.05 & 1022.8 & \textbf{50.43} & 511.4 & 48.83 & \cellcolor{green!15}25.7 & 49.99 & 40.6 \\
 &  & RT-J & 51.08 & 1011.0 & \textbf{54.89} & 506.1 & 52.68 & \cellcolor{green!15}25.2 & 53.26 & 34.0 \\
 &  & Griffin & \textbf{50.97} & 14699.2 & 50.93 & 7491.2 & 47.42 & \cellcolor{green!15}1193.7 & 50.42 & 2798.8 \\
\midrule
\multirow{3}{*}{Mean} & \multirow{3}{*}{(14 tasks)}
 & RT & 69.98 & 915.6 & 66.42 & \cellcolor{green!15}457.9 & \textbf{71.46} & 546.6 & 71.44 & 578.2 \\
 &  & RT-J & 71.72 & 912.6 & 67.63 & \cellcolor{green!15}456.8 & 71.64 & 517.7 & \textbf{72.37} & 554.3 \\
 &  & Griffin & \textbf{65.29} & 31503.9 & 63.18 & \cellcolor{green!15}15895.7 & 64.50 & 16319.1 & 65.08 & 16552.5 \\
\midrule
\multirow{3}{*}{$\Delta$} & \multirow{3}{*}{vs.\ baseline}
 & RT & --- & --- & -3.56 & -50.0\% & +1.48 & -40.3\% & +1.46 & -36.9\% \\
 &  & RT-J & --- & --- & -4.10 & -49.9\% & -0.08 & -43.3\% & +0.64 & -39.3\% \\
 &  & Griffin & --- & --- & -2.11 & -49.5\% & -0.78 & -48.2\% & -0.21 & -47.5\% \\
\bottomrule
\end{tabularx}
\end{table*}

\begin{table*}[p]
\centering
\small
\caption{Main results: regression tasks ($R^2$~\%, higher is better). Used in \Cref{fig:main-reg}. For each task and model, we report test accuracy and number of tokens with the baseline (random sampling), naive pruning (full-budget sampling, then half the resulting sample discarded at random), STEER with Claude-based ranking, and STEER with GPT-based ranking. Within each row, bold indicates the highest test score and the highlighted cell indicates the lowest token count (ties are broken in favor of one non-baseline condition). The Mean block at the bottom of the table reports the per-model average across all 10 tasks. The $\Delta$ block reports the change relative to the baseline (point difference for test accuracy, and percentage reduction for tokens).}
\label{tab:appendix-main-reg}
\begin{tabularx}{\textwidth}{ccl YY YY YY YY}
\toprule
\multirow{2}{*}{Database} & \multirow{2}{*}{Task} & \multirow{2}{*}{RFM} & \multicolumn{2}{c}{Baseline} & \multicolumn{2}{c}{Naive Pruning} & \multicolumn{2}{c}{STEER (Claude)} & \multicolumn{2}{c}{STEER (GPT)} \\
\cmidrule(lr){4-5} \cmidrule(lr){6-7} \cmidrule(lr){8-9} \cmidrule(lr){10-11}
& & & Test $\uparrow$ & Tokens $\downarrow$ & Test $\uparrow$ & Tokens $\downarrow$ & Test $\uparrow$ & Tokens $\downarrow$ & Test $\uparrow$ & Tokens $\downarrow$ \\
\midrule
\multirow{3}{*}{rel-amazon} & \multirow{3}{*}{item-ltv} & RT & \textbf{4.59} & 957.2 & 3.03 & 478.7 & 4.26 & \cellcolor{green!15}203.5 & 4.26 & 203.5 \\
 &  & RT-J & 59.31 & 951.5 & 24.29 & 476.2 & \textbf{60.17} & \cellcolor{green!15}108.4 & 60.17 & 108.4 \\
 &  & Griffin & \textbf{15.70} & 10175.6 & 13.27 & \cellcolor{green!15}5167.8 & 15.38 & 8079.4 & 15.56 & 7241.0 \\
\midrule
\multirow{3}{*}{rel-amazon} & \multirow{3}{*}{user-ltv} & RT & 15.20 & 1023.2 & 3.42 & 511.6 & \textbf{26.26} & \cellcolor{green!15}100.8 & 25.29 & 101.0 \\
 &  & RT-J & \textbf{61.52} & 1021.3 & 11.26 & 511.1 & 57.19 & 86.9 & 56.01 & \cellcolor{green!15}86.7 \\
 &  & Griffin & \textbf{-0.49} & 5906.3 & -7.51 & \cellcolor{green!15}3033.2 & -5.45 & 4048.7 & -1.10 & 5377.0 \\
\midrule
\multirow{3}{*}{rel-arxiv} & \multirow{3}{*}{author-publication} & RT & 22.77 & 740.5 & 23.00 & \cellcolor{green!15}370.3 & \textbf{24.75} & 491.4 & 23.99 & 387.1 \\
 &  & RT-J & 18.46 & 728.8 & 7.72 & 364.8 & 33.45 & 474.4 & \textbf{35.36} & \cellcolor{green!15}356.6 \\
 &  & Griffin & -1.67 & 2140.3 & \textbf{1.99} & \cellcolor{green!15}1126.1 & -1.67 & 2140.3 & -1.06 & 1937.5 \\
\midrule
\multirow{3}{*}{rel-avito} & \multirow{3}{*}{ad-ctr} & RT & 4.19 & 1024.0 & 1.91 & \cellcolor{green!15}512.0 & 5.88 & 999.4 & \textbf{7.70} & 1024.0 \\
 &  & RT-J & \textbf{2.44} & 1020.8 & 1.67 & \cellcolor{green!15}511.0 & -3.02 & 994.6 & -2.77 & 1018.6 \\
 &  & Griffin & 2.39 & 16593.3 & 0.51 & 8379.8 & \textbf{2.96} & 9025.7 & 2.42 & \cellcolor{green!15}6927.3 \\
\midrule
\multirow{3}{*}{rel-f1} & \multirow{3}{*}{driver-position} & RT & 45.31 & 982.6 & 35.16 & \cellcolor{green!15}491.3 & 44.69 & 982.6 & \textbf{46.90} & 982.6 \\
 &  & RT-J & 55.40 & 978.9 & 39.16 & \cellcolor{green!15}490.0 & 55.34 & 978.4 & \textbf{55.55} & 978.4 \\
 &  & Griffin & -2.74 & 36402.7 & -2.65 & \cellcolor{green!15}18322.9 & -2.59 & 28600.6 & \textbf{-2.40} & 25318.9 \\
\midrule
\multirow{3}{*}{rel-hm} & \multirow{3}{*}{item-sales} & RT & 21.19 & 751.5 & 2.27 & 375.8 & \textbf{23.74} & \cellcolor{green!15}313.4 & 21.62 & 640.5 \\
 &  & RT-J & 44.12 & 729.3 & 32.16 & 365.0 & \textbf{44.30} & \cellcolor{green!15}186.6 & 42.81 & 622.8 \\
 &  & Griffin & \textbf{5.54} & 30219.9 & 5.37 & \cellcolor{green!15}15307.8 & 5.48 & 25540.4 & 5.48 & 26478.5 \\
\midrule
\multirow{3}{*}{rel-ratebeer} & \multirow{3}{*}{user-rating-count} & RT & 12.91 & 1024.0 & 13.67 & 512.0 & \textbf{19.47} & \cellcolor{green!15}420.2 & 18.21 & 432.9 \\
 &  & RT-J & 40.83 & 1013.0 & 41.83 & 507.0 & \textbf{45.06} & \cellcolor{green!15}272.6 & 44.66 & 289.0 \\
 &  & Griffin & -5.74 & 32905.8 & \textbf{-5.64} & \cellcolor{green!15}16727.2 & -5.73 & 23975.0 & -5.69 & 19543.1 \\
\midrule
\multirow{3}{*}{rel-stack} & \multirow{3}{*}{post-votes} & RT & 28.75 & 1003.0 & 22.37 & 501.5 & \textbf{29.18} & \cellcolor{green!15}413.5 & 28.22 & 677.9 \\
 &  & RT-J & \textbf{86.08} & 989.1 & 58.84 & 495.1 & 85.25 & \cellcolor{green!15}139.5 & 81.05 & 590.5 \\
 &  & Griffin & 17.75 & 24390.1 & 13.09 & 12309.2 & \textbf{18.03} & \cellcolor{green!15}9879.3 & 17.50 & 10707.1 \\
\midrule
\multirow{3}{*}{rel-trial} & \multirow{3}{*}{site-success} & RT & 14.78 & 1015.3 & 8.52 & \cellcolor{green!15}507.6 & \textbf{34.83} & 944.9 & 22.89 & 787.5 \\
 &  & RT-J & 43.54 & 924.8 & 32.96 & \cellcolor{green!15}463.0 & \textbf{46.66} & 794.8 & 33.44 & 699.6 \\
 &  & Griffin & 1.29 & 6860.1 & 0.40 & \cellcolor{green!15}3575.4 & 1.29 & 6860.1 & \textbf{1.33} & 6099.5 \\
\midrule
\multirow{3}{*}{rel-trial} & \multirow{3}{*}{study-adverse} & RT & 3.57 & 1024.0 & 1.86 & 512.0 & -0.92 & \cellcolor{green!15}21.7 & \textbf{3.97} & 77.2 \\
 &  & RT-J & -7.37 & 1014.1 & \textbf{-6.40} & 507.6 & -12.19 & \cellcolor{green!15}22.0 & -7.18 & 45.2 \\
 &  & Griffin & -2.52 & 11558.5 & -2.29 & 5923.7 & -2.93 & \cellcolor{green!15}947.0 & \textbf{-1.92} & 1944.4 \\
\midrule
\multirow{3}{*}{Mean} & \multirow{3}{*}{(10 tasks)}
 & RT & 17.33 & 954.5 & 11.52 & \cellcolor{green!15}477.3 & \textbf{21.21} & 489.1 & 20.31 & 531.4 \\
 &  & RT-J & 40.43 & 937.2 & 24.35 & 469.1 & \textbf{41.22} & \cellcolor{green!15}405.8 & 39.91 & 479.6 \\
 &  & Griffin & 2.95 & 17715.3 & 1.65 & \cellcolor{green!15}8987.3 & 2.48 & 11909.7 & \textbf{3.01} & 11157.4 \\
\midrule
\multirow{3}{*}{$\Delta$} & \multirow{3}{*}{vs.\ baseline}
 & RT & --- & --- & -5.80 & -50.0\% & +3.89 & -48.8\% & +2.98 & -44.3\% \\
 &  & RT-J & --- & --- & -16.09 & -49.9\% & +0.79 & -56.7\% & -0.52 & -48.8\% \\
 &  & Griffin & --- & --- & -1.30 & -49.3\% & -0.47 & -32.8\% & +0.06 & -37.0\% \\
\bottomrule
\end{tabularx}
\end{table*}

\begin{table*}[t]
\centering
\small
\caption{Direction of traversal ablation: classification tasks (AUROC~\%, higher is better). Used in \Cref{fig:direction-cls}. Ablation results use only the RT and RT-J models and the Claude LLM. For each task and model, we report test accuracy and number of tokens with the baseline (random sampling), \PtoF{}-only traversal, \FtoP{}-only traversal, and the STEER traversal filter that combines \PtoF{} and \FtoP{}. Within each row, bold indicates the highest test score and the highlighted cell indicates the lowest token count (ties are broken in favor of one non-baseline condition). The Mean block at the bottom of the table reports the per-model average across all 14 tasks. The $\Delta$ block reports the change relative to the baseline (point difference for test accuracy, and percentage reduction for tokens).}
\label{tab:appendix-direction-cls}
\begin{tabularx}{\textwidth}{ccl YY YY YY YY}
\toprule
\multirow{2}{*}{Database} & \multirow{2}{*}{Task} & \multirow{2}{*}{RFM} & \multicolumn{2}{c}{Baseline} & \multicolumn{2}{c}{\PtoF{} only} & \multicolumn{2}{c}{\FtoP{} only} & \multicolumn{2}{c}{STEER} \\
\cmidrule(lr){4-5} \cmidrule(lr){6-7} \cmidrule(lr){8-9} \cmidrule(lr){10-11}
& & & Test $\uparrow$ & Tokens $\downarrow$ & Test $\uparrow$ & Tokens $\downarrow$ & Test $\uparrow$ & Tokens $\downarrow$ & Test $\uparrow$ & Tokens $\downarrow$ \\
\midrule
\multirow{2}{*}{rel-amazon} & \multirow{2}{*}{item-churn} & RT & 68.20 & 1023.2 & \textbf{71.25} & 259.9 & 71.06 & \cellcolor{green!15}230.2 & 71.06 & 230.2 \\
 &  & RT-J & 72.26 & 1018.9 & \textbf{74.59} & 137.8 & 74.29 & \cellcolor{green!15}118.7 & 74.21 & 118.7 \\
\midrule
\multirow{2}{*}{rel-amazon} & \multirow{2}{*}{user-churn} & RT & 64.14 & 1024.0 & \textbf{65.41} & 124.0 & 63.19 & 738.6 & 65.29 & \cellcolor{green!15}82.5 \\
 &  & RT-J & 65.45 & 1021.3 & 65.30 & 105.5 & 63.50 & 742.2 & \textbf{65.53} & \cellcolor{green!15}68.1 \\
\midrule
\multirow{2}{*}{rel-arxiv} & \multirow{2}{*}{paper-citation} & RT & 79.85 & 989.6 & \textbf{80.30} & 937.0 & 79.92 & \cellcolor{green!15}856.8 & 79.99 & 863.2 \\
 &  & RT-J & 80.54 & 997.3 & \textbf{80.78} & 941.8 & 79.81 & 873.8 & 79.99 & \cellcolor{green!15}862.7 \\
\midrule
\multirow{2}{*}{rel-avito} & \multirow{2}{*}{user-clicks} & RT & 51.92 & 613.6 & 57.31 & 545.9 & 61.56 & 497.0 & \textbf{63.80} & \cellcolor{green!15}487.5 \\
 &  & RT-J & \textbf{49.64} & 612.2 & 42.50 & 549.6 & 43.51 & 476.0 & 40.52 & \cellcolor{green!15}463.6 \\
\midrule
\multirow{2}{*}{rel-avito} & \multirow{2}{*}{user-visits} & RT & 55.98 & 706.7 & 58.95 & 418.4 & 56.08 & 676.9 & \textbf{59.47} & \cellcolor{green!15}374.8 \\
 &  & RT-J & 63.77 & 731.7 & 64.44 & 210.3 & 64.05 & 703.5 & \textbf{64.94} & \cellcolor{green!15}150.7 \\
\midrule
\multirow{2}{*}{rel-f1} & \multirow{2}{*}{driver-dnf} & RT & 78.55 & 1005.2 & 80.26 & 1002.3 & 80.30 & 1000.9 & \textbf{81.40} & \cellcolor{green!15}999.4 \\
 &  & RT-J & \textbf{78.50} & 1001.5 & 77.37 & 1000.8 & 78.32 & 997.2 & 77.58 & \cellcolor{green!15}995.3 \\
\midrule
\multirow{2}{*}{rel-f1} & \multirow{2}{*}{driver-top3} & RT & 89.26 & 1000.2 & 90.49 & \cellcolor{green!15}1000.2 & 89.72 & 1000.2 & \textbf{90.70} & 1000.2 \\
 &  & RT-J & 89.23 & 996.6 & \textbf{89.46} & \cellcolor{green!15}996.5 & 89.45 & 996.6 & 89.15 & 996.8 \\
\midrule
\multirow{2}{*}{rel-hm} & \multirow{2}{*}{user-churn} & RT & 63.80 & 1024.0 & 64.48 & 595.0 & \textbf{65.75} & 918.3 & 65.40 & \cellcolor{green!15}283.3 \\
 &  & RT-J & \textbf{63.63} & 1020.3 & 61.50 & 543.0 & 61.61 & 910.4 & 61.49 & \cellcolor{green!15}197.6 \\
\midrule
\multirow{2}{*}{rel-ratebeer} & \multirow{2}{*}{beer-rating-churn} & RT & 68.25 & 1024.0 & 67.19 & 1024.0 & 70.59 & 1022.7 & \textbf{70.93} & \cellcolor{green!15}1005.8 \\
 &  & RT-J & 66.98 & 1011.7 & 71.05 & 1015.0 & 67.12 & 1009.5 & \textbf{74.82} & \cellcolor{green!15}976.7 \\
\midrule
\multirow{2}{*}{rel-ratebeer} & \multirow{2}{*}{brewer-dormant} & RT & \textbf{71.79} & 1024.0 & 70.32 & 691.4 & 70.56 & 981.8 & 69.54 & \cellcolor{green!15}621.4 \\
 &  & RT-J & 61.66 & 1007.8 & 62.08 & 695.7 & \textbf{62.51} & 965.5 & 61.83 & \cellcolor{green!15}631.8 \\
\midrule
\multirow{2}{*}{rel-ratebeer} & \multirow{2}{*}{user-rating-churn} & RT & 73.38 & 1024.0 & 71.19 & 991.6 & \textbf{74.31} & 913.4 & 73.02 & \cellcolor{green!15}887.3 \\
 &  & RT-J & 84.29 & 1013.1 & \textbf{86.02} & 991.6 & 85.82 & 913.5 & 85.72 & \cellcolor{green!15}898.2 \\
\midrule
\multirow{2}{*}{rel-stack} & \multirow{2}{*}{user-badge} & RT & \textbf{81.69} & 355.6 & 78.94 & 256.8 & 79.14 & 303.2 & 75.89 & \cellcolor{green!15}134.2 \\
 &  & RT-J & 89.53 & 357.8 & 89.65 & 256.3 & \textbf{90.38} & 311.8 & 87.01 & \cellcolor{green!15}125.0 \\
\midrule
\multirow{2}{*}{rel-stack} & \multirow{2}{*}{user-engagement} & RT & 82.83 & 981.5 & \textbf{85.70} & 765.3 & 81.04 & 919.0 & 85.15 & \cellcolor{green!15}656.8 \\
 &  & RT-J & 87.59 & 975.0 & \textbf{88.41} & 833.8 & 87.31 & 934.4 & 87.52 & \cellcolor{green!15}737.3 \\
\midrule
\multirow{2}{*}{rel-trial} & \multirow{2}{*}{study-outcome} & RT & 50.05 & 1022.8 & 48.98 & 26.4 & \textbf{53.53} & 899.4 & 48.83 & \cellcolor{green!15}25.7 \\
 &  & RT-J & 51.08 & 1011.0 & 52.53 & 25.8 & 48.05 & 756.6 & \textbf{52.68} & \cellcolor{green!15}25.2 \\
\midrule
\multirow{2}{*}{Mean} & \multirow{2}{*}{(14 tasks)}
 & RT & 69.98 & 915.6 & 70.77 & 617.0 & 71.20 & 782.7 & \textbf{71.46} & \cellcolor{green!15}546.6 \\
 &  & RT-J & 71.72 & 912.6 & \textbf{71.83} & 593.1 & 71.12 & 765.0 & 71.64 & \cellcolor{green!15}517.7 \\
\midrule
\multirow{2}{*}{$\Delta$} & \multirow{2}{*}{vs.\ baseline}
 & RT & --- & --- & +0.79 & -32.6\% & +1.22 & -14.5\% & +1.48 & -40.3\% \\
 &  & RT-J & --- & --- & +0.11 & -35.0\% & -0.60 & -16.2\% & -0.08 & -43.3\% \\
\bottomrule
\end{tabularx}
\end{table*}

\begin{table*}[t]
\centering
\small
\caption{Direction of traversal ablation: regression tasks ($R^2$~\%, higher is better). Used in \Cref{fig:direction-reg}. Ablation results use only the RT and RT-J models and the Claude LLM. For each task and model, we report test accuracy and number of tokens with the baseline (random sampling), \PtoF{}-only traversal, \FtoP{}-only traversal, and the STEER traversal filter that combines \PtoF{} and \FtoP{}. Within each row, bold indicates the highest test score and the highlighted cell indicates the lowest token count (ties are broken in favor of one non-baseline condition). The Mean block at the bottom of the table reports the per-model average across all 10 tasks. The $\Delta$ block reports the change relative to the baseline (point difference for test accuracy, and percentage reduction for tokens).}
\label{tab:appendix-direction-reg}
\begin{tabularx}{\textwidth}{ccl YY YY YY YY}
\toprule
\multirow{2}{*}{Database} & \multirow{2}{*}{Task} & \multirow{2}{*}{RFM} & \multicolumn{2}{c}{Baseline} & \multicolumn{2}{c}{\PtoF{} only} & \multicolumn{2}{c}{\FtoP{} only} & \multicolumn{2}{c}{STEER} \\
\cmidrule(lr){4-5} \cmidrule(lr){6-7} \cmidrule(lr){8-9} \cmidrule(lr){10-11}
& & & Test $\uparrow$ & Tokens $\downarrow$ & Test $\uparrow$ & Tokens $\downarrow$ & Test $\uparrow$ & Tokens $\downarrow$ & Test $\uparrow$ & Tokens $\downarrow$ \\
\midrule
\multirow{2}{*}{rel-amazon} & \multirow{2}{*}{item-ltv} & RT & \textbf{4.59} & 957.2 & 4.46 & 954.1 & 4.26 & \cellcolor{green!15}203.5 & 4.26 & 203.5 \\
 &  & RT-J & 59.31 & 951.5 & 58.89 & 945.4 & 60.00 & \cellcolor{green!15}108.4 & \textbf{60.17} & 108.4 \\
\midrule
\multirow{2}{*}{rel-amazon} & \multirow{2}{*}{user-ltv} & RT & 15.20 & 1023.2 & 24.13 & 119.5 & 25.99 & 851.9 & \textbf{26.26} & \cellcolor{green!15}100.8 \\
 &  & RT-J & \textbf{61.52} & 1021.3 & 53.86 & 105.5 & 53.66 & 602.3 & 57.19 & \cellcolor{green!15}86.9 \\
\midrule
\multirow{2}{*}{rel-arxiv} & \multirow{2}{*}{author-publication} & RT & 22.77 & 740.5 & \textbf{24.75} & \cellcolor{green!15}491.4 & 22.64 & 680.6 & 24.75 & 491.4 \\
 &  & RT-J & 18.46 & 728.8 & 30.50 & \cellcolor{green!15}453.3 & 21.46 & 684.0 & \textbf{33.45} & 474.4 \\
\midrule
\multirow{2}{*}{rel-avito} & \multirow{2}{*}{ad-ctr} & RT & 4.19 & 1024.0 & 4.56 & 1024.0 & \textbf{6.90} & 1000.9 & 5.88 & \cellcolor{green!15}999.4 \\
 &  & RT-J & \textbf{2.44} & 1020.8 & -0.28 & 1020.2 & 2.09 & 995.2 & -3.02 & \cellcolor{green!15}994.6 \\
\midrule
\multirow{2}{*}{rel-f1} & \multirow{2}{*}{driver-position} & RT & \textbf{45.31} & 982.6 & 45.26 & \cellcolor{green!15}982.6 & 44.72 & 982.6 & 44.69 & 982.6 \\
 &  & RT-J & 55.40 & 978.9 & 55.78 & 978.6 & \textbf{55.84} & 978.8 & 55.34 & \cellcolor{green!15}978.4 \\
\midrule
\multirow{2}{*}{rel-hm} & \multirow{2}{*}{item-sales} & RT & 21.19 & 751.5 & 21.98 & 409.2 & \textbf{23.74} & \cellcolor{green!15}313.4 & 23.74 & 313.4 \\
 &  & RT-J & 44.12 & 729.3 & \textbf{44.49} & 250.2 & 44.40 & \cellcolor{green!15}186.6 & 44.30 & 186.6 \\
\midrule
\multirow{2}{*}{rel-ratebeer} & \multirow{2}{*}{user-rating-count} & RT & 12.91 & 1024.0 & 12.61 & 613.5 & \textbf{27.15} & 719.7 & 19.47 & \cellcolor{green!15}420.2 \\
 &  & RT-J & 40.83 & 1013.0 & 38.75 & 613.1 & \textbf{54.89} & 706.6 & 45.06 & \cellcolor{green!15}272.6 \\
\midrule
\multirow{2}{*}{rel-stack} & \multirow{2}{*}{post-votes} & RT & 28.75 & 1003.0 & 24.57 & 868.6 & 28.07 & 708.6 & \textbf{29.18} & \cellcolor{green!15}413.5 \\
 &  & RT-J & \textbf{86.08} & 989.1 & 85.22 & 581.4 & 85.44 & 714.2 & 85.25 & \cellcolor{green!15}139.5 \\
\midrule
\multirow{2}{*}{rel-trial} & \multirow{2}{*}{site-success} & RT & 14.78 & 1015.3 & 33.45 & 986.0 & 18.06 & 977.9 & \textbf{34.83} & \cellcolor{green!15}944.9 \\
 &  & RT-J & 43.54 & 924.8 & \textbf{47.24} & 826.7 & 45.19 & 837.2 & 46.66 & \cellcolor{green!15}794.8 \\
\midrule
\multirow{2}{*}{rel-trial} & \multirow{2}{*}{study-adverse} & RT & \textbf{3.57} & 1024.0 & -0.82 & 21.9 & 3.44 & 879.0 & -0.92 & \cellcolor{green!15}21.7 \\
 &  & RT-J & -7.37 & 1014.1 & -11.90 & 22.2 & \textbf{-6.84} & 814.2 & -12.19 & \cellcolor{green!15}22.0 \\
\midrule
\multirow{2}{*}{Mean} & \multirow{2}{*}{(10 tasks)}
 & RT & 17.33 & 954.5 & 19.49 & 647.1 & 20.50 & 731.8 & \textbf{21.21} & \cellcolor{green!15}489.1 \\
 &  & RT-J & 40.43 & 937.2 & 40.25 & 579.7 & \textbf{41.61} & 662.8 & 41.22 & \cellcolor{green!15}405.8 \\
\midrule
\multirow{2}{*}{$\Delta$} & \multirow{2}{*}{vs.\ baseline}
 & RT & --- & --- & +2.17 & -32.2\% & +3.17 & -23.3\% & +3.89 & -48.8\% \\
 &  & RT-J & --- & --- & -0.18 & -38.1\% & +1.18 & -29.3\% & +0.79 & -56.7\% \\
\bottomrule
\end{tabularx}
\end{table*}

\begin{table*}[t]
\centering
\small
\caption{Prompting strategy ablation: classification tasks (AUROC~\%, higher is better). Used in \Cref{tab:prompting-summary}. Ablation results use only the RT and RT-J models. For each task and model, we report test accuracy with the baseline (random sampling). We also report simple prompting and STEER's smart prompting with four LLM variants: Claude and GPT with high and default reasoning effort. Within each simple/STEER pair, bold indicates the higher accuracy (ties favor STEER). The Mean block at the bottom of the table reports the per-model average across all 14 tasks. The $\Delta$ block reports the change relative to the baseline (point difference).}
\label{tab:appendix-prompting-cls}
\begin{tabularx}{\textwidth}{cclY YY YY YY YY}
\toprule
\multirow{2}{*}{Database} & \multirow{2}{*}{Task} & \multirow{2}{*}{RFM} & \multirow{2}{*}{Base.} & \multicolumn{2}{c}{Claude High} & \multicolumn{2}{c}{Claude Default} & \multicolumn{2}{c}{GPT High} & \multicolumn{2}{c}{GPT Default} \\
\cmidrule(lr){5-6} \cmidrule(lr){7-8} \cmidrule(lr){9-10} \cmidrule(lr){11-12}
& & & & Simple & STEER & Simple & STEER & Simple & STEER & Simple & STEER \\
\midrule
\multirow{2}{*}{rel-amazon} & \multirow{2}{*}{item-churn} & RT & 68.20 & \textbf{71.25} & 71.06 & \textbf{71.25} & 71.07 & 71.06 & \textbf{71.07} & 71.06 & \textbf{71.06} \\
 &  & RT-J & 72.26 & \textbf{74.76} & 74.21 & \textbf{74.76} & 74.41 & 74.34 & \textbf{74.41} & \textbf{74.34} & 74.21 \\
\midrule
\multirow{2}{*}{rel-amazon} & \multirow{2}{*}{user-churn} & RT & 64.14 & \textbf{65.41} & 65.29 & \textbf{65.41} & 65.29 & 64.55 & \textbf{64.55} & \textbf{65.41} & 65.29 \\
 &  & RT-J & 65.45 & \textbf{65.54} & 65.53 & \textbf{65.54} & 65.53 & 65.05 & \textbf{65.05} & \textbf{65.54} & 65.53 \\
\midrule
\multirow{2}{*}{rel-arxiv} & \multirow{2}{*}{paper-citation} & RT & 79.85 & 79.85 & \textbf{79.99} & \textbf{79.85} & 79.83 & 74.21 & \textbf{79.39} & 74.21 & \textbf{79.12} \\
 &  & RT-J & 80.54 & \textbf{80.51} & 79.99 & \textbf{80.51} & 80.27 & 75.01 & \textbf{80.81} & 75.01 & \textbf{79.43} \\
\midrule
\multirow{2}{*}{rel-avito} & \multirow{2}{*}{user-clicks} & RT & 51.92 & \textbf{64.79} & 63.80 & 58.30 & \textbf{64.60} & 51.92 & \textbf{63.02} & 58.30 & \textbf{58.36} \\
 &  & RT-J & 49.64 & 34.01 & \textbf{40.52} & \textbf{41.76} & 40.84 & \textbf{45.69} & 45.30 & 41.76 & \textbf{44.42} \\
\midrule
\multirow{2}{*}{rel-avito} & \multirow{2}{*}{user-visits} & RT & 55.98 & 58.81 & \textbf{59.47} & \textbf{58.97} & 58.33 & 55.98 & \textbf{59.57} & 55.98 & \textbf{57.14} \\
 &  & RT-J & 63.77 & 64.76 & \textbf{64.94} & \textbf{64.78} & 64.35 & 63.26 & \textbf{65.07} & 63.26 & \textbf{65.53} \\
\midrule
\multirow{2}{*}{rel-f1} & \multirow{2}{*}{driver-dnf} & RT & 78.55 & 78.74 & \textbf{81.40} & 78.74 & \textbf{80.64} & 78.74 & \textbf{80.62} & 78.74 & \textbf{79.96} \\
 &  & RT-J & 78.50 & \textbf{79.19} & 77.58 & \textbf{79.19} & 78.06 & \textbf{79.19} & 78.44 & \textbf{79.19} & 78.00 \\
\midrule
\multirow{2}{*}{rel-f1} & \multirow{2}{*}{driver-top3} & RT & 89.26 & 89.36 & \textbf{90.70} & 89.68 & \textbf{90.62} & \textbf{89.58} & 89.35 & 89.48 & \textbf{90.05} \\
 &  & RT-J & 89.23 & \textbf{90.07} & 89.15 & \textbf{89.96} & 89.18 & \textbf{89.41} & 89.18 & \textbf{89.54} & 89.06 \\
\midrule
\multirow{2}{*}{rel-hm} & \multirow{2}{*}{user-churn} & RT & 63.80 & 64.48 & \textbf{65.40} & 64.48 & \textbf{65.40} & 65.26 & \textbf{65.40} & 64.48 & \textbf{65.40} \\
 &  & RT-J & 63.63 & 61.30 & \textbf{61.49} & 61.30 & \textbf{61.49} & 60.96 & \textbf{61.49} & 61.30 & \textbf{61.49} \\
\midrule
\multirow{2}{*}{rel-ratebeer} & \multirow{2}{*}{beer-rating-churn} & RT & 68.25 & 67.42 & \textbf{70.93} & 63.83 & \textbf{68.36} & 64.08 & \textbf{70.25} & 67.61 & \textbf{69.98} \\
 &  & RT-J & 66.98 & 68.06 & \textbf{74.82} & 67.94 & \textbf{71.77} & 68.02 & \textbf{75.75} & 68.07 & \textbf{74.15} \\
\midrule
\multirow{2}{*}{rel-ratebeer} & \multirow{2}{*}{brewer-dormant} & RT & 71.79 & \textbf{70.07} & 69.54 & \textbf{70.86} & 69.11 & \textbf{70.01} & 70.01 & \textbf{70.07} & 69.12 \\
 &  & RT-J & 61.66 & \textbf{62.52} & 61.83 & \textbf{62.49} & 61.39 & 62.59 & \textbf{63.96} & \textbf{62.52} & 62.28 \\
\midrule
\multirow{2}{*}{rel-ratebeer} & \multirow{2}{*}{user-rating-churn} & RT & 73.38 & \textbf{73.32} & 73.02 & 73.34 & \textbf{73.82} & \textbf{81.51} & 72.37 & \textbf{81.51} & 78.40 \\
 &  & RT-J & 84.29 & 84.22 & \textbf{85.72} & 83.64 & \textbf{84.56} & 84.07 & \textbf{86.10} & \textbf{84.07} & 81.86 \\
\midrule
\multirow{2}{*}{rel-stack} & \multirow{2}{*}{user-badge} & RT & 81.69 & \textbf{81.49} & 75.89 & \textbf{81.69} & 75.89 & 70.58 & \textbf{78.77} & 71.35 & \textbf{77.60} \\
 &  & RT-J & 89.53 & \textbf{89.78} & 87.01 & 86.52 & \textbf{87.01} & 83.62 & \textbf{87.50} & 84.27 & \textbf{87.82} \\
\midrule
\multirow{2}{*}{rel-stack} & \multirow{2}{*}{user-engagement} & RT & 82.83 & 82.83 & \textbf{85.15} & 82.57 & \textbf{85.00} & \textbf{88.39} & 85.80 & 83.22 & \textbf{87.47} \\
 &  & RT-J & 87.59 & 87.03 & \textbf{87.52} & \textbf{90.59} & 90.29 & \textbf{87.42} & 86.81 & \textbf{92.12} & 90.44 \\
\midrule
\multirow{2}{*}{rel-trial} & \multirow{2}{*}{study-outcome} & RT & 50.05 & \textbf{49.15} & 48.83 & 49.15 & \textbf{50.24} & 49.15 & \textbf{49.99} & \textbf{49.15} & 48.83 \\
 &  & RT-J & 51.08 & 49.83 & \textbf{52.68} & 49.83 & \textbf{52.08} & 49.83 & \textbf{53.26} & 49.83 & \textbf{52.68} \\
\midrule
\multirow{2}{*}{Mean} & \multirow{2}{*}{(14 tasks)}
 & RT & 69.98 & 71.21 & \textbf{71.46} & 70.58 & \textbf{71.30} & 69.64 & \textbf{71.44} & 70.04 & \textbf{71.27} \\
 &  & RT-J & 71.72 & 70.83 & \textbf{71.64} & 71.34 & \textbf{71.52} & 70.60 & \textbf{72.37} & 70.77 & \textbf{71.92} \\
\midrule
\multirow{2}{*}{$\Delta$} & \multirow{2}{*}{vs.\ baseline}
 & RT & --- & +1.23 & +1.48 & +0.60 & +1.32 & -0.33 & +1.46 & +0.06 & +1.29 \\
 &  & RT-J & --- & -0.90 & -0.08 & -0.38 & -0.21 & -1.12 & +0.64 & -0.95 & +0.20 \\
\bottomrule
\end{tabularx}
\end{table*}

\begin{table*}[t]
\centering
\small
\caption{Prompting strategy ablation: regression tasks ($R^2$~\%, higher is better). Used in \Cref{tab:prompting-summary}. Ablation results use only the RT and RT-J models. For each task and model, we report test accuracy with the baseline (random sampling). We also report simple prompting and STEER's smart prompting with four LLM variants: Claude and GPT with high and default reasoning effort. Within each simple/STEER pair, bold indicates the higher accuracy (ties favor STEER). The Mean block at the bottom of the table reports the per-model average across all 10 tasks. The $\Delta$ block reports the change relative to the baseline (point difference).}
\label{tab:appendix-prompting-reg}
\begin{tabularx}{\textwidth}{cclY YY YY YY YY}
\toprule
\multirow{2}{*}{Database} & \multirow{2}{*}{Task} & \multirow{2}{*}{RFM} & \multirow{2}{*}{Base.} & \multicolumn{2}{c}{Claude High} & \multicolumn{2}{c}{Claude Default} & \multicolumn{2}{c}{GPT High} & \multicolumn{2}{c}{GPT Default} \\
\cmidrule(lr){5-6} \cmidrule(lr){7-8} \cmidrule(lr){9-10} \cmidrule(lr){11-12}
& & & & Simple & STEER & Simple & STEER & Simple & STEER & Simple & STEER \\
\midrule
\multirow{2}{*}{rel-amazon} & \multirow{2}{*}{item-ltv} & RT & 4.59 & 4.00 & \textbf{4.26} & 4.00 & \textbf{4.26} & 4.00 & \textbf{4.26} & \textbf{4.26} & 4.22 \\
 &  & RT-J & 59.31 & 59.05 & \textbf{60.17} & 59.05 & \textbf{60.17} & 59.05 & \textbf{60.17} & \textbf{60.17} & 59.84 \\
\midrule
\multirow{2}{*}{rel-amazon} & \multirow{2}{*}{user-ltv} & RT & 15.20 & 24.13 & \textbf{26.26} & 24.13 & \textbf{25.29} & \textbf{25.73} & 25.29 & \textbf{25.73} & 23.90 \\
 &  & RT-J & 61.52 & 54.60 & \textbf{57.19} & 54.60 & \textbf{56.01} & 54.41 & \textbf{56.01} & 54.41 & \textbf{55.79} \\
\midrule
\multirow{2}{*}{rel-arxiv} & \multirow{2}{*}{author-publication} & RT & 22.77 & 22.99 & \textbf{24.75} & 22.80 & \textbf{24.30} & 23.14 & \textbf{23.99} & 22.57 & \textbf{23.15} \\
 &  & RT-J & 18.46 & 14.56 & \textbf{33.45} & 14.57 & \textbf{31.08} & 14.64 & \textbf{35.36} & 14.76 & \textbf{26.00} \\
\midrule
\multirow{2}{*}{rel-avito} & \multirow{2}{*}{ad-ctr} & RT & 4.19 & \textbf{8.07} & 5.88 & \textbf{8.07} & 3.82 & 4.48 & \textbf{7.70} & 4.48 & \textbf{6.03} \\
 &  & RT-J & 2.44 & \textbf{1.84} & -3.02 & \textbf{1.84} & -3.40 & \textbf{1.87} & -2.77 & \textbf{1.87} & -1.92 \\
\midrule
\multirow{2}{*}{rel-f1} & \multirow{2}{*}{driver-position} & RT & 45.31 & \textbf{45.14} & 44.69 & 45.38 & \textbf{45.72} & 45.33 & \textbf{46.90} & 45.18 & \textbf{45.25} \\
 &  & RT-J & 55.40 & \textbf{56.29} & 55.34 & \textbf{57.24} & 55.67 & \textbf{56.63} & 55.55 & \textbf{56.78} & 55.35 \\
\midrule
\multirow{2}{*}{rel-hm} & \multirow{2}{*}{item-sales} & RT & 21.19 & 21.98 & \textbf{23.74} & 21.98 & \textbf{23.74} & \textbf{21.98} & 21.62 & \textbf{21.98} & 21.22 \\
 &  & RT-J & 44.12 & \textbf{44.42} & 44.30 & \textbf{44.42} & 44.30 & \textbf{44.42} & 42.81 & \textbf{44.42} & 43.49 \\
\midrule
\multirow{2}{*}{rel-ratebeer} & \multirow{2}{*}{user-rating-count} & RT & 12.91 & 15.41 & \textbf{19.47} & 20.66 & \textbf{26.18} & 8.51 & \textbf{18.21} & 22.47 & \textbf{26.10} \\
 &  & RT-J & 40.83 & 32.44 & \textbf{45.06} & 32.33 & \textbf{44.84} & 30.91 & \textbf{44.66} & \textbf{46.40} & 44.90 \\
\midrule
\multirow{2}{*}{rel-stack} & \multirow{2}{*}{post-votes} & RT & 28.75 & 27.31 & \textbf{29.18} & \textbf{27.31} & 19.04 & 23.05 & \textbf{28.22} & 23.05 & \textbf{27.74} \\
 &  & RT-J & 86.08 & \textbf{85.68} & 85.25 & \textbf{85.68} & 84.83 & \textbf{84.59} & 81.05 & \textbf{84.59} & 84.09 \\
\midrule
\multirow{2}{*}{rel-trial} & \multirow{2}{*}{site-success} & RT & 14.78 & 1.81 & \textbf{34.83} & 1.84 & \textbf{37.40} & 2.20 & \textbf{22.89} & -2.59 & \textbf{22.49} \\
 &  & RT-J & 43.54 & -0.69 & \textbf{46.66} & -0.82 & \textbf{44.84} & -1.10 & \textbf{33.44} & -0.59 & \textbf{33.30} \\
\midrule
\multirow{2}{*}{rel-trial} & \multirow{2}{*}{study-adverse} & RT & 3.57 & \textbf{5.05} & -0.92 & 4.30 & \textbf{5.37} & \textbf{5.05} & 3.97 & 4.30 & \textbf{5.86} \\
 &  & RT-J & -7.37 & \textbf{-6.97} & -12.19 & \textbf{-6.75} & -7.99 & \textbf{-6.97} & -7.18 & \textbf{-6.75} & -8.07 \\
\midrule
\multirow{2}{*}{Mean} & \multirow{2}{*}{(10 tasks)}
 & RT & 17.33 & 17.59 & \textbf{21.21} & 18.05 & \textbf{21.51} & 16.35 & \textbf{20.31} & 17.14 & \textbf{20.60} \\
 &  & RT-J & 40.43 & 34.12 & \textbf{41.22} & 34.22 & \textbf{41.03} & 33.85 & \textbf{39.91} & 35.61 & \textbf{39.28} \\
\midrule
\multirow{2}{*}{$\Delta$} & \multirow{2}{*}{vs.\ baseline}
 & RT & --- & +0.26 & +3.89 & +0.72 & +4.18 & -0.98 & +2.98 & -0.18 & +3.27 \\
 &  & RT-J & --- & -6.31 & +0.79 & -6.22 & +0.60 & -6.59 & -0.52 & -4.83 & -1.16 \\
\bottomrule
\end{tabularx}
\end{table*}

\begin{table*}[t]
\centering
\small
\caption{Column filter ablation: classification tasks (AUROC~\%, higher is better). Used in \Cref{tab:colfilter-summary}. Ablation results use only the RT and RT-J models and the Claude LLM.  For each task and model, we report test accuracy and number of tokens with the baseline (random sampling), STEER without column filter, and STEER with column filter. Within each row, bold indicates the highest test score and the highlighted cell indicates the lowest token count (ties are broken in favor of one non-baseline condition). The Mean block at the bottom of the table reports the per-model average across all 14 tasks. The $\Delta$ block reports the change relative to the baseline (point difference for test accuracy, and percentage reduction for tokens).}
\label{tab:appendix-colfilter-cls}
\begin{tabularx}{\textwidth}{ccl YY YY YY}
\toprule
\multirow{2}{*}{Database} & \multirow{2}{*}{Task} & \multirow{2}{*}{RFM} & \multicolumn{2}{c}{Baseline} & \multicolumn{2}{c}{STEER} & \multicolumn{2}{c}{STEER + Col.\ Filter} \\
\cmidrule(lr){4-5} \cmidrule(lr){6-7} \cmidrule(lr){8-9}
& & & Test $\uparrow$ & Tokens $\downarrow$ & Test $\uparrow$ & Tokens $\downarrow$ & Test $\uparrow$ & Tokens $\downarrow$ \\
\midrule
\multirow{2}{*}{rel-amazon} & \multirow{2}{*}{item-churn} & RT & 68.20 & 1023.2 & \textbf{71.06} & 230.2 & 69.98 & \cellcolor{green!15}91.3 \\
 &  & RT-J & 72.26 & 1018.9 & \textbf{74.21} & 118.7 & 72.67 & \cellcolor{green!15}50.8 \\
\midrule
\multirow{2}{*}{rel-amazon} & \multirow{2}{*}{user-churn} & RT & 64.14 & 1024.0 & \textbf{65.29} & 82.5 & 63.75 & \cellcolor{green!15}35.5 \\
 &  & RT-J & 65.45 & 1021.3 & \textbf{65.53} & 68.1 & 65.38 & \cellcolor{green!15}30.2 \\
\midrule
\multirow{2}{*}{rel-arxiv} & \multirow{2}{*}{paper-citation} & RT & 79.85 & 989.6 & \textbf{79.99} & 863.2 & 78.37 & \cellcolor{green!15}862.0 \\
 &  & RT-J & \textbf{80.54} & 997.3 & 79.99 & \cellcolor{green!15}862.7 & 79.34 & 862.8 \\
\midrule
\multirow{2}{*}{rel-avito} & \multirow{2}{*}{user-clicks} & RT & 51.92 & 613.6 & \textbf{63.80} & 487.5 & 61.20 & \cellcolor{green!15}482.2 \\
 &  & RT-J & \textbf{49.64} & 612.2 & 40.52 & 463.6 & 43.74 & \cellcolor{green!15}458.8 \\
\midrule
\multirow{2}{*}{rel-avito} & \multirow{2}{*}{user-visits} & RT & 55.98 & 706.7 & \textbf{59.47} & 374.8 & 56.55 & \cellcolor{green!15}201.5 \\
 &  & RT-J & 63.77 & 731.7 & \textbf{64.94} & 150.7 & 62.47 & \cellcolor{green!15}59.3 \\
\midrule
\multirow{2}{*}{rel-f1} & \multirow{2}{*}{driver-dnf} & RT & 78.55 & 1005.2 & \textbf{81.40} & 999.4 & 81.27 & \cellcolor{green!15}999.3 \\
 &  & RT-J & 78.50 & 1001.5 & 77.58 & \cellcolor{green!15}995.3 & \textbf{80.32} & 997.6 \\
\midrule
\multirow{2}{*}{rel-f1} & \multirow{2}{*}{driver-top3} & RT & 89.26 & 1000.2 & \textbf{90.70} & 1000.2 & 90.52 & \cellcolor{green!15}1000.1 \\
 &  & RT-J & 89.23 & \cellcolor{green!15}996.6 & 89.15 & 996.8 & \textbf{89.75} & 999.2 \\
\midrule
\multirow{2}{*}{rel-hm} & \multirow{2}{*}{user-churn} & RT & 63.80 & 1024.0 & 65.40 & 283.3 & \textbf{65.77} & \cellcolor{green!15}62.4 \\
 &  & RT-J & \textbf{63.63} & 1020.3 & 61.49 & 197.6 & 62.33 & \cellcolor{green!15}43.0 \\
\midrule
\multirow{2}{*}{rel-ratebeer} & \multirow{2}{*}{beer-rating-churn} & RT & 68.25 & 1024.0 & 70.93 & 1005.8 & \textbf{71.71} & \cellcolor{green!15}1001.5 \\
 &  & RT-J & 66.98 & 1011.7 & 74.82 & 976.7 & \textbf{84.53} & \cellcolor{green!15}972.7 \\
\midrule
\multirow{2}{*}{rel-ratebeer} & \multirow{2}{*}{brewer-dormant} & RT & \textbf{71.79} & 1024.0 & 69.54 & 621.4 & 71.19 & \cellcolor{green!15}142.9 \\
 &  & RT-J & 61.66 & 1007.8 & \textbf{61.83} & 631.8 & 60.30 & \cellcolor{green!15}78.9 \\
\midrule
\multirow{2}{*}{rel-ratebeer} & \multirow{2}{*}{user-rating-churn} & RT & 73.38 & 1024.0 & 73.02 & 887.3 & \textbf{79.56} & \cellcolor{green!15}883.9 \\
 &  & RT-J & 84.29 & 1013.1 & 85.72 & \cellcolor{green!15}898.2 & \textbf{89.04} & 902.3 \\
\midrule
\multirow{2}{*}{rel-stack} & \multirow{2}{*}{user-badge} & RT & \textbf{81.69} & 355.6 & 75.89 & 134.2 & 72.24 & \cellcolor{green!15}126.3 \\
 &  & RT-J & \textbf{89.53} & 357.8 & 87.01 & 125.0 & 80.93 & \cellcolor{green!15}110.5 \\
\midrule
\multirow{2}{*}{rel-stack} & \multirow{2}{*}{user-engagement} & RT & 82.83 & 981.5 & \textbf{85.15} & 656.8 & 78.09 & \cellcolor{green!15}646.8 \\
 &  & RT-J & \textbf{87.59} & 975.0 & 87.52 & 737.3 & 86.09 & \cellcolor{green!15}735.0 \\
\midrule
\multirow{2}{*}{rel-trial} & \multirow{2}{*}{study-outcome} & RT & 50.05 & 1022.8 & 48.83 & 25.7 & \textbf{51.33} & \cellcolor{green!15}8.2 \\
 &  & RT-J & 51.08 & 1011.0 & 52.68 & 25.2 & \textbf{53.91} & \cellcolor{green!15}7.7 \\
\midrule
\multirow{2}{*}{Mean} & \multirow{2}{*}{(14 tasks)}
 & RT & 69.98 & 915.6 & \textbf{71.46} & 546.6 & 70.82 & \cellcolor{green!15}467.4 \\
 &  & RT-J & 71.72 & 912.6 & 71.64 & 517.7 & \textbf{72.20} & \cellcolor{green!15}450.6 \\
\midrule
\multirow{2}{*}{$\Delta$} & \multirow{2}{*}{vs.\ baseline}
 & RT & --- & --- & +1.48 & -40.3\% & +0.84 & -48.9\% \\
 &  & RT-J & --- & --- & -0.08 & -43.3\% & +0.48 & -50.6\% \\
\bottomrule
\end{tabularx}
\end{table*}

\begin{table*}[t]
\centering
\small
\caption{Column filter ablation: regression tasks ($R^2$~\%, higher is better). Used in \Cref{tab:colfilter-summary}. Ablation results use only the RT and RT-J models and the Claude LLM.  For each task and model, we report test accuracy and number of tokens with the baseline (random sampling), STEER without column filter, and STEER with column filter. Within each row, bold indicates the highest test score and the highlighted cell indicates the lowest token count (ties are broken in favor of one non-baseline condition). The Mean block at the bottom of the table reports the per-model average across all 10 tasks. The $\Delta$ block reports the change relative to the baseline (point difference for test accuracy, and percentage reduction for tokens).}
\label{tab:appendix-colfilter-reg}
\begin{tabularx}{\textwidth}{ccl YY YY YY}
\toprule
\multirow{2}{*}{Database} & \multirow{2}{*}{Task} & \multirow{2}{*}{RFM} & \multicolumn{2}{c}{Baseline} & \multicolumn{2}{c}{STEER} & \multicolumn{2}{c}{STEER + Col.\ Filter} \\
\cmidrule(lr){4-5} \cmidrule(lr){6-7} \cmidrule(lr){8-9}
& & & Test $\uparrow$ & Tokens $\downarrow$ & Test $\uparrow$ & Tokens $\downarrow$ & Test $\uparrow$ & Tokens $\downarrow$ \\
\midrule
\multirow{2}{*}{rel-amazon} & \multirow{2}{*}{item-ltv} & RT & 4.59 & 957.2 & 4.26 & 203.5 & \textbf{33.25} & \cellcolor{green!15}42.9 \\
 &  & RT-J & 59.31 & 951.5 & \textbf{60.17} & 108.4 & 59.64 & \cellcolor{green!15}30.2 \\
\midrule
\multirow{2}{*}{rel-amazon} & \multirow{2}{*}{user-ltv} & RT & 15.20 & 1023.2 & 26.26 & 100.8 & \textbf{26.95} & \cellcolor{green!15}50.1 \\
 &  & RT-J & \textbf{61.52} & 1021.3 & 57.19 & 86.9 & 22.33 & \cellcolor{green!15}41.7 \\
\midrule
\multirow{2}{*}{rel-arxiv} & \multirow{2}{*}{author-publication} & RT & 22.77 & 740.5 & \textbf{24.75} & 491.4 & 11.63 & \cellcolor{green!15}480.0 \\
 &  & RT-J & 18.46 & 728.8 & \textbf{33.45} & 474.4 & 19.43 & \cellcolor{green!15}461.4 \\
\midrule
\multirow{2}{*}{rel-avito} & \multirow{2}{*}{ad-ctr} & RT & 4.19 & 1024.0 & 5.88 & 999.4 & \textbf{7.02} & \cellcolor{green!15}999.3 \\
 &  & RT-J & \textbf{2.44} & 1020.8 & -3.02 & \cellcolor{green!15}994.6 & -6.72 & 995.4 \\
\midrule
\multirow{2}{*}{rel-f1} & \multirow{2}{*}{driver-position} & RT & \textbf{45.31} & 982.6 & 44.69 & 982.6 & 44.07 & \cellcolor{green!15}982.5 \\
 &  & RT-J & \textbf{55.40} & 978.9 & 55.34 & \cellcolor{green!15}978.4 & 54.59 & 980.6 \\
\midrule
\multirow{2}{*}{rel-hm} & \multirow{2}{*}{item-sales} & RT & 21.19 & 751.5 & \textbf{23.74} & 313.4 & 23.34 & \cellcolor{green!15}245.5 \\
 &  & RT-J & 44.12 & 729.3 & \textbf{44.30} & 186.6 & 40.63 & \cellcolor{green!15}142.3 \\
\midrule
\multirow{2}{*}{rel-ratebeer} & \multirow{2}{*}{user-rating-count} & RT & 12.91 & 1024.0 & 19.47 & 420.2 & \textbf{50.34} & \cellcolor{green!15}30.3 \\
 &  & RT-J & 40.83 & 1013.0 & 45.06 & 272.6 & \textbf{52.65} & \cellcolor{green!15}22.3 \\
\midrule
\multirow{2}{*}{rel-stack} & \multirow{2}{*}{post-votes} & RT & 28.75 & 1003.0 & 29.18 & 413.5 & \textbf{29.80} & \cellcolor{green!15}392.2 \\
 &  & RT-J & \textbf{86.08} & 989.1 & 85.25 & 139.5 & 82.42 & \cellcolor{green!15}98.0 \\
\midrule
\multirow{2}{*}{rel-trial} & \multirow{2}{*}{site-success} & RT & 14.78 & 1015.3 & 34.83 & 944.9 & \textbf{42.74} & \cellcolor{green!15}864.6 \\
 &  & RT-J & 43.54 & 924.8 & 46.66 & 794.8 & \textbf{63.62} & \cellcolor{green!15}762.1 \\
\midrule
\multirow{2}{*}{rel-trial} & \multirow{2}{*}{study-adverse} & RT & \textbf{3.57} & 1024.0 & -0.92 & 21.7 & -3.13 & \cellcolor{green!15}5.0 \\
 &  & RT-J & \textbf{-7.37} & 1014.1 & -12.19 & 22.0 & -7.48 & \cellcolor{green!15}5.0 \\
\midrule
\multirow{2}{*}{Mean} & \multirow{2}{*}{(10 tasks)}
 & RT & 17.33 & 954.5 & 21.21 & 489.1 & \textbf{26.60} & \cellcolor{green!15}409.2 \\
 &  & RT-J & 40.43 & 937.2 & \textbf{41.22} & 405.8 & 38.11 & \cellcolor{green!15}353.9 \\
\midrule
\multirow{2}{*}{$\Delta$} & \multirow{2}{*}{vs.\ baseline}
 & RT & --- & --- & +3.89 & -48.8\% & +9.28 & -57.1\% \\
 &  & RT-J & --- & --- & +0.79 & -56.7\% & -2.32 & -62.2\% \\
\bottomrule
\end{tabularx}
\end{table*}

\end{document}